\documentclass[journal]{IEEEtran}

\usepackage{graphicx} 
\usepackage{todonotes}
\usepackage{booktabs}
\usepackage{graphicx} 
\usepackage{amsmath}
\usepackage{amssymb}
\usepackage{cite}
\usepackage{url}
\usepackage{hyperref}
\usepackage{color}

\usepackage[ruled,vlined]{algorithm2e}

\begin{document}

\title{Buried Fiber-Optic Geolocalization with Distributed Acoustic Sensing}

\author{Khen~Cohen, Natanel~Nissan, Ofir~Nissan, and Ariel~Lellouch
\thanks{Khen Cohen is with The School of Physics and Astronomy, Tel-Aviv University, Tel-Aviv 69978, Israel (e-mail: khencohen@mail.tau.ac.il).}
\thanks{Natanel Nissan and Ofir Nissan are with the School of Electrical and Computer Engineering, Tel-Aviv University, Tel-Aviv 69978, Israel.}
\thanks{Ariel Lellouch is with the Department of Geophysics, Tel-Aviv University, Tel-Aviv 69978, Israel.}}
\maketitle

\begin{abstract}
We present a scalable method for geolocalizing buried fiber-optic cables using Distributed Acoustic Sensing (DAS) and traffic-induced quasi-static seismic signals. Assuming access to one end of the fiber, the method fuses DAS measurements with vehicle trajectories obtained from either video tracking or vehicle-mounted GPS. The fiber geometry is estimated by minimizing the mismatch between the measured and physics-based synthetic strain-rate maps. The framework combines a matched-filter initialization with neural-network-based trajectory optimization, enabling robust convergence under realistic noise and trajectory-uncertainty conditions.
Simulation and field experiments demonstrate sub-meter localization accuracy, often on the order of tens of centimeters, and strong agreement with manual calibration by tap-testing. This approach provides a practical tool for mapping poorly documented underground fiber infrastructure and for supporting urban sensing applications.
\end{abstract}


\section{Introduction}
\label{sec:intro}

The urban environment contains dense underground networks of fiber-optic cables. However, utility records often suffer from significant positional inaccuracies. A comparative study that examined utility records against Subsurface Utility Engineering (SUE) investigations found that only 32\% of utilities aligned within 2 feet ($\approx 0.6m$), while approximately 21\% deviated by more than 20 feet ($\approx 6.1m$) \cite{ASCE3802}. As a consequence, frequent damage to existing but poorly mapped fiber-optic infrastructure during construction and excavation leads to IT service disruptions, repair costs, and project delays. In dense urban settings, sub-meter mapping accuracy is therefore critical to reduce accidental damage during construction. The challenge is compounded by the physical properties of fiber-optic cables: because both the fiber and its protective casing are typically non-conductive, conventional underground utility detection methods, such as electromagnetic (EM) detection and ground-penetrating radar (GPR), often cannot reliably locate fiber-optic infrastructure. However, the existence of manholes or optical cabinets, visible at the surface, can be leveraged to constrain fiber routes. 

Over the past decade, Distributed Fiber Optic Sensing (DFOS) technologies, particularly Distributed Acoustic Sensing (DAS), have shown that existing telecommunication fiber can be transformed into high-resolution seismic monitoring arrays \cite{dasantennas, fosexample, pendao2022opticalFOSSurvey1, das4monitorIfrast}. Many studies have extended this capability to urban environments and demonstrated the value of vehicle-signal extraction for traffic monitoring and management \cite{vehicleextraction, wang2022urban,arthur2024urban,liu2023telecomtm,Cohen2025FiberOptic}. The main advantage of DAS over traditional seismic sensors, which is the meter-scale spatial channel resolution it offers, is fully realized only when the fiber path is precisely known \cite{martin2017continuous}. A key prerequisite for urban DAS applications is thus the accurate geolocalization of the fiber itself. In addition to the need for the location of the fiber, the DAS measurements are directional along the fiber axis and are therefore inherently tied to fiber geometry \cite{vesselDetectionLocalization, Cohen2025FiberOptic, wang2025trafficflowspeedmonitoring, xie2024intelligenttrafficmonitoringdistributed, Kou24VehicleDetection}. Hence, accurate fiber geolocalization is also crucial for correct interpretation of the directional sensitivity of the DAS measurements. Consequently, precise mapping of underground fiber cables, especially legacy installations with incomplete documentation, remains a central barrier to large-scale DAS deployments \cite{Wang2020, SurveyFOS3}.

Early fiber-localization methods relied on tap tests, in which seismic signals were generated at known surface positions (e.g., with weight drops or sledgehammers) and then matched to peak DAS channel responses in strain or strain rate \cite{liu2020new}. More recent approaches automated this idea by synchronizing vehicle-mounted GPS with DAS data, achieving meter-level localization along the roadway direction, while cross-road (perpendicular) accuracy remains limited {\cite{biondi2023geolocalization}}. A complementary strategy uses access points (manholes) as localization anchors. In DAS data, manholes can often be identified because the spooled slack fiber inside them exhibits a seismic response different from that of straight buried segments, including under distant earthquake wavefields \cite{bukharin2023ambient,lindsey2017eqwavefields,ajo2019darkfiber}. Although manhole detection provides useful anchor points, it is generally insufficient on its own to recover the full fiber trajectory at sub-meter precision \cite{bukharin2023ambient}.

{Fiber geolocalization using raw DAS measurements introduces additional challenges.} The recordings are frequently contaminated by multiple noise sources such as optical fading, environmental interference, and instrument noise, all of which degrade the signal-to-noise ratio (SNR) \cite{feng2025zsn2n, zhao2022denoiser}. The phase estimation of large strains, such as those induced by direct tap-test hits above the cable, is unstable and depends on the specific phase unwrapping algorithm \cite{hartog2017introduction}. Extracting useful information from vehicle signals is complicated by near-field effects and variable source characteristics, which can introduce substantial uncertainty if not modeled carefully \cite{ding2025robust}. Recent work has highlighted the need for advanced deep-learning pipelines to mitigate these artifacts, as well as the computational burden imposed by continuous DAS acquisition and massive data volumes \cite{chen2025compression}. Therefore, even when DAS data are available, deriving reliable geometric information remains nontrivial and motivates the development of robust, automated localization methods.

\begin{figure*}
    \centering
    \includegraphics[width=1.0\linewidth]{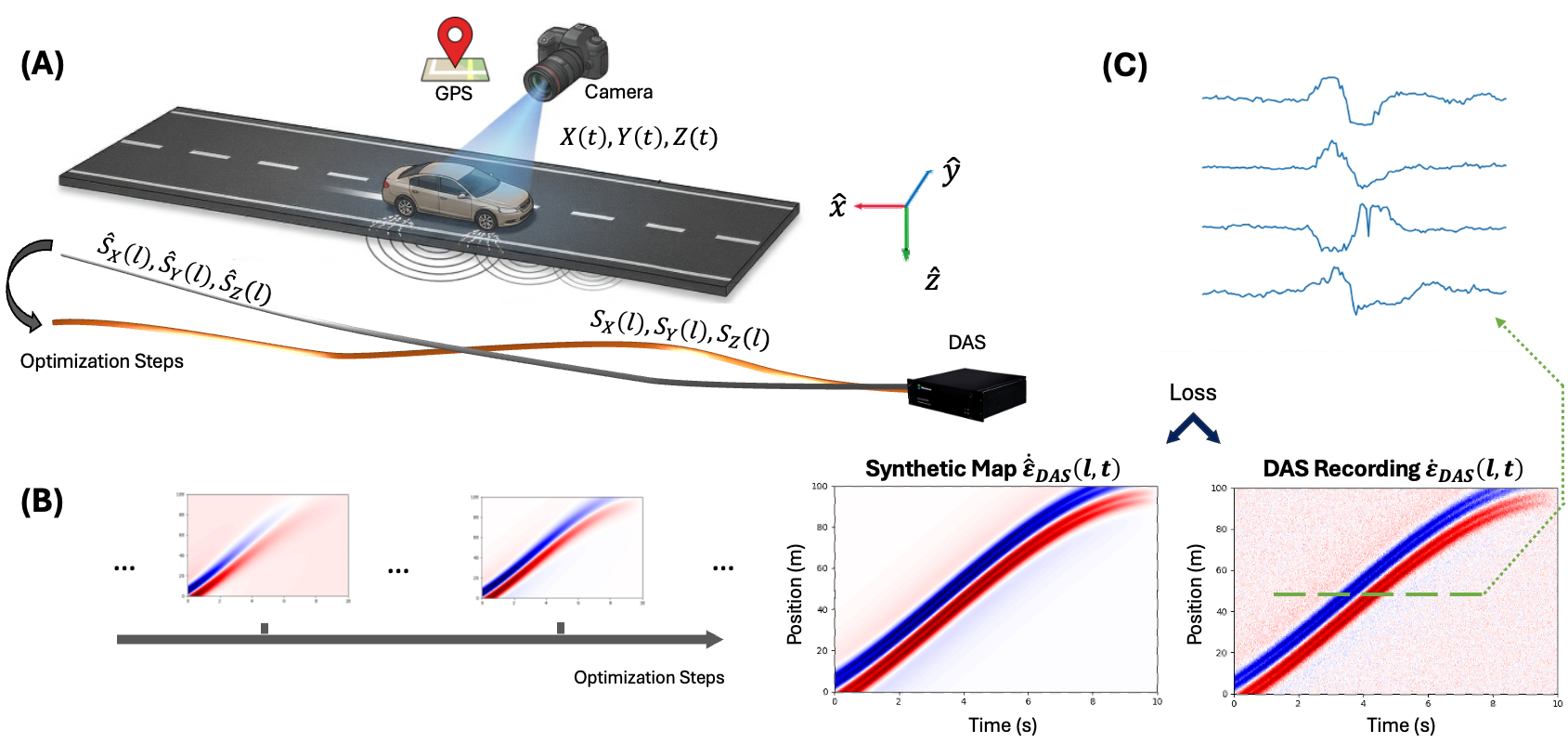}
    \caption{Diagram of the fiber localization method.
    (A) A driving car is continuously recorded by a camera or GPS, while a nearby DAS system measures seismic waves via the optical fiber (orange). The gray fiber represents an initial guess of the fiber’s location.
    (B) During the optimization process, the synthetic strain rate generated from the estimated fiber path converges toward the real strain rate measured by the DAS.
    (C) Experimental DAS recordings of strain rate versus time at different locations in the presence of a moving vehicle.}
    \label{fig:mainDiagram}
\end{figure*}

In this study, we present an automatic and scalable method for sub-meter localization of underground fiber utilities. As in prior work, we leverage ambient traffic noise. However, whereas traditional approaches often use only the maximum DAS amplitude at the channel nearest to a vehicle as a location proxy, our method models the full spatiotemporal DAS response induced by moving vehicles and estimates the fiber trajectory that best fits the measurements. Figure \ref{fig:mainDiagram} provides an overview of the workflow. We show that, with vehicle trajectory information obtained either from vehicle-mounted GPS or from camera-based tracking, the method achieves sub-meter localization accuracy under realistic conditions. The remainder of this paper is organized as follows: Section II presents the mathematical and physical background. Section III introduces the localization framework, including model-based sensitivity analysis and a two-stage optimization strategy. Section IV reports simulation results across multiple operating conditions. Section V presents two experiments on the same street (camera-based and GPS-based), and Section VI concludes with limitations and future directions.

\section{Background}
\label{sec:background}
\subsection{Fiber Location and Coordinate System}

We consider a buried optical fiber of total length $L_{\text{tot}}$, accurately measurable from OTDR readings. We define a global Cartesian coordinate system with orthonormal basis vectors $\left(\hat{\mathbf{x}}, \hat{\mathbf{y}}, \hat{\mathbf{z}}\right)$. The spatial location of the fiber in this coordinate frame is described by a 3D curve parameterized by the arc-length coordinate $l$ (measured from a reference point along the fiber):
\begin{equation}
    \mathbf{S}(l) = \big(x(l),\; y(l),\; z(l)\big) \ .
\end{equation}

As the DAS interrogator outputs measurements at discrete intervals along the fiber, we represent its layout as a sequence of $N$ uniformly spaced points along its path:
\begin{equation}
    \mathbf{S}_n = \big(x_n, y_n, z_n\big), \quad n = 1, \dots, N \ .
\end{equation}
With $N$ determined by the fiber length and spatial sampling interval $\Delta x$, i.e.,
\begin{equation}
    N = \frac{L_{\text{tot}}}{\Delta x} \ ,
\end{equation}
we refer to the spatial sampling interval as a \emph{segment}. In this study, the optical acquisition parameters yield $\Delta x \approx 1\,\mathrm{m}$. We emphasize that this value is not the same as spatial resolution, because DAS measurements are averaged over a finite gauge length \cite{hartog2017introduction}, which is typically on the order of 3-10 m in urban monitoring settings. 

\subsection{Distance Metric Between Possible Fiber Trajectories}
To quantify the discrepancy between the true and predicted trajectories, denoted by $\mathbf{S}_n$ and $\mathbf{\hat{S}}_m$, respectively, we use the \emph{Hausdorff distance} between the corresponding point sets:
\begin{align} \label{eq:hausd}
    d_H(\mathbf{S}_n, \mathbf{\hat{S}}_m) = 
    \max\bigg\{ & 
        \sup_{\mathbf{x} \in \mathbf{S}_n} \inf_{\mathbf{y} \in \mathbf{\hat{S}}_m} \|\mathbf{x} - \mathbf{y}\|,
        \; \\ \nonumber
        & \sup_{\mathbf{y} \in \mathbf{\hat{S}}_m} \inf_{\mathbf{x} \in \mathbf{S}_n} \|\mathbf{x} - \mathbf{y}\|
    \bigg\} \ .
\end{align}
This metric is well suited for trajectory comparison because it quantifies the maximum mismatch between the two curves, ensuring that localized deviations are not overlooked. The Hausdorff distance is zero only when the two trajectories coincide as point sets. In addition, it naturally handles trajectories with different numbers of sampled points. All trajectory-distance evaluations in this paper use this definition.

\subsection{Camera Model}
\label{sub:camera_model}

We adopt the camera modeling framework described in \cite{szeliski2022computer}. Each event $i$ captured by the camera at time $t$ is associated with a pixel location $(x_i(t), y_i(t))$ in the image plane. The relationship between pixel coordinates and 3D world coordinates is given by the camera projection matrix $P$:
\begin{equation}
    \mathbf{p}_i \propto P \, \mathbf{X}_i \ ,
\end{equation}
where $\mathbf{p}_i$ is the homogeneous pixel coordinate and $\mathbf{X}_i$ is the corresponding homogeneous 3D point. In general, this transformation is not directly invertible, as all points lying along the same camera ray are projected to the same pixel. 

To resolve this ambiguity, we assume a locally planar and horizontal ground surface. The 3D position of each observed point is then determined by intersecting its camera ray with the ground plane. This assumption enables a direct mapping from pixel coordinates to 3D object locations in the global coordinate system. Throughout this paper, we assume that vehicles move along the $x$ axis, the $y$ axis represents lateral (cross-road) offset, and the $z$ axis represents depth.

\subsection{Vehicle-Induced {Strain Field} in Fiber Coordinates}
\label{subsec:seismic_wavefield_fiber}

Moving vehicles generate a {strain field} with two principal components \cite{vandenende, yuan2023, yuan2020near}. The first is a quasi-static component (frequencies $<1$~Hz), arising from elastic ground deformation under the vehicle load. The second is a dynamic component, typically in the $\sim$2-20~Hz range, dominated by Rayleigh-type surface waves excited by road-wheel interactions. Although recent studies have used rigorous 3D forward modeling to characterize the full wavefield generated by heavy moving sources (e.g., high-speed trains) \cite{wang2024forward}, our geometric-localization framework for urban settings focuses on quasi-static deformation. Prior studies \cite{vandenende, yuan2020near} show that the Flamant-Boussinesq solution for a point load is an effective approximation of the recorded signal. The displacement field $u_{x}(x, y, z)$ along the x-axis, produced by a vertical point load $F$ at the origin, is given for a straight fiber by:
\begin{equation}
    u_{x}(x, y, z) = \frac{F}{4\pi G} \frac{x}{r^2} \left( \frac{z}{r} + \frac{2\nu - 1}{1 + \frac{z}{r}} \right) \ ,
    \label{eq:eq_u}
\end{equation}
where $(x, y, z)$ are coordinates in the global frame, $r = \sqrt{x^2 + y^2 + z^2}$ is the Euclidean distance from the load point to the measurement point, and $G$ and $\nu$ are the shear modulus and Poisson's ratio of the subsurface, assumed homogeneous and known. However, in the case of a general fiber layout, the displacements are measured along the trajectory rather than a specific axis. In addition, the coordinates $(x, y, z)$ should be expressed in the \emph{fiber coordinate system} rather than in the global frame. Appendix~\ref{app:FiberSystem} provides the corrected formulation in fiber coordinates. While we supply it for completeness, we find that this correction has a negligible effect on the results reported in this study. 

DAS systems, as opposed to traditional sensors, do not measure point strain or strain rate. Instead, they measure the accumulated strain/strain-rate along a gauge length, which is typically in the range of 3-10 m in urban monitoring settings. The directional strain rate at position $l$ along the fiber and time $t$ is thus equivalent to the differential directional displacement rate along the gauge length:
\begin{equation}
    \dot{\varepsilon}_{\mathrm{DAS}}(l, t) = \frac{1}{{L_g}} \left[ \dot{u}_{l}\left(l + \frac{{L_g}}{2}, t \right) - \dot{u}_{l}\left(l - \frac{{L_g}}{2}, t \right) \right] \ ,
    \label{eq:e_dot_das}
\end{equation}
where $\dot{u}_{l}$ is the temporal derivative of displacement along the fiber direction{, and $L_g$ denotes the gauge length}.
In practice, the load from a moving vehicle can be represented as a superposition of point loads at the wheel locations. The total strain-rate signal recorded by the DAS is then obtained by summing the contributions from all loads with appropriate time-shifts. For the quasi-static part of the {strain field}, the response is assumed immediate and no temporal propagation effects need to be taken into account. 

\section{Fiber Localization Method}
\label{sec:method}
In this section, we present the methodology used to localize the optical fiber. We begin with a sensitivity analysis based primarily on Eq.~\ref{eq:e_dot_das}. We then introduce a two-stage localization framework: (i) an initialization step based on matched filtering of the strain-rate map, and (ii) a neural-network-based optimization stage that estimates fiber coordinates. Together, these components enable high-precision fiber localization.

An accurate model for a vehicle load requires representing each wheel as a separate point load. However, our analysis shows that this model improved overall accuracy only by a few centimeters on average, and only in some of the tests, for our data acquisition configuration with a 10-m gauge length. In addition, estimating axle distance may not be straightforward for large vehicles. Therefore, we model vehicles as a point load. The vehicle trajectory is obtained from a camera tracking algorithm or vehicle-mounted GPS, as will be discussed in \ref{sec:experiment}.

The subsurface parameters, shear modulus $G$ and Poisson ratio $\nu$, are assumed spatially constant. The absolute value of $G$ is not needed as the synthetic map is normalized (see \ref{sub:synthetic_map_gen}). However, $\nu$ significantly influences the shape of the signal. Previous studies have also attempted to estimate it directly from quasi-static measurements \cite{yuan2020near}. The spatial homogeneity assumption may thus break down over long distances or in areas of sharp lithological changes. Therefore, we apply our method to road intervals of up to approximately 100~m to limit such changes { while any longer fiber can be built using such shorter fiber segments, as we show in Appendix \ref{app:FiberScalability}}.

\subsection{Sensitivity Analysis}
\label{sub:sensAnalysis}

We begin with a simplified scenario, in which a single point load moves approximately parallel and at a constant horizontal distance to the optical fiber. A gauge length of 10 m is used for the DAS strain-rate generation. In this setting, the first step of localization is estimating the fiber coordinate along the road direction (the $x$ axis). Under the quasi-static approximation, for any fixed time $t$, the displacement profile along the fiber reaches its maximum magnitude at the point closest to the moving load. Equivalently, the corresponding \textit{strain-rate} signal crosses zero at that point (hereafter, the \textit{crossing point}). Thus, when the vehicle is observed at position $x(t)$, the channel index associated with the strain-rate zero crossing, $l(t)$, identifies the nearest fiber location to the vehicle. The $x$ coordinate of that fiber point is then estimated from the known $x(t)$ at the same time.

\begin{figure}
    \centering
    \includegraphics[width=1.0\linewidth]{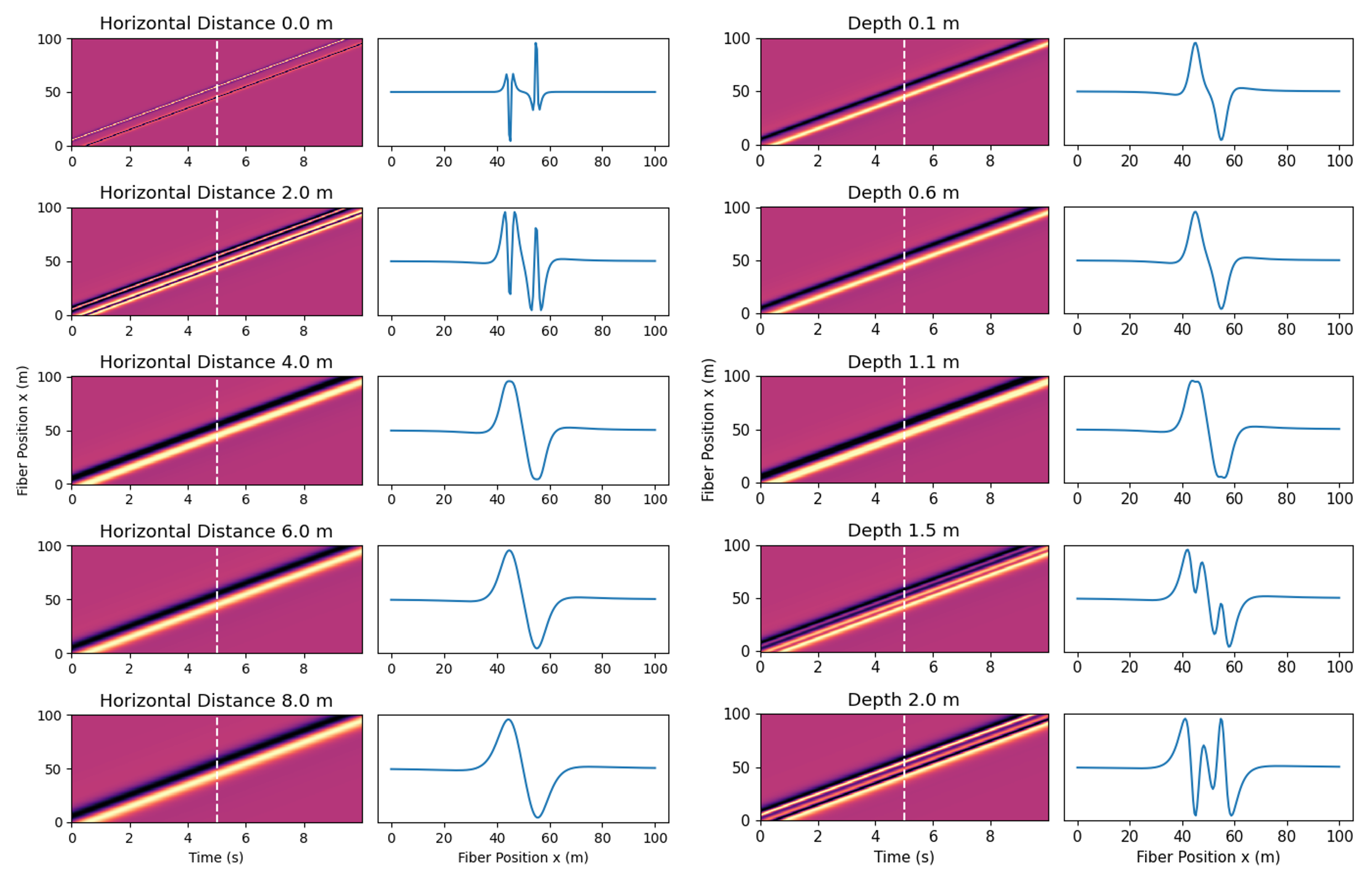}
    \caption{Spatio-temporal strain rates of a point object moving at constant speed for different geometrical settings. The spatial-only strain rate, extracted at t=5 s, is also shown. Left - different horizontal distances for a fixed depth of 1 m. The spatial strain-rate signature broadens with horizontal distance, and stabilizes around 4 m. Right - different fiber depths for a fixed horizontal distance of 4 m. Increase of burial depth complexifies recorded signal. }
    \label{fig:widthVsDistance}
\end{figure}

\begin{figure}
    \centering
    \includegraphics[width=1.0\linewidth]{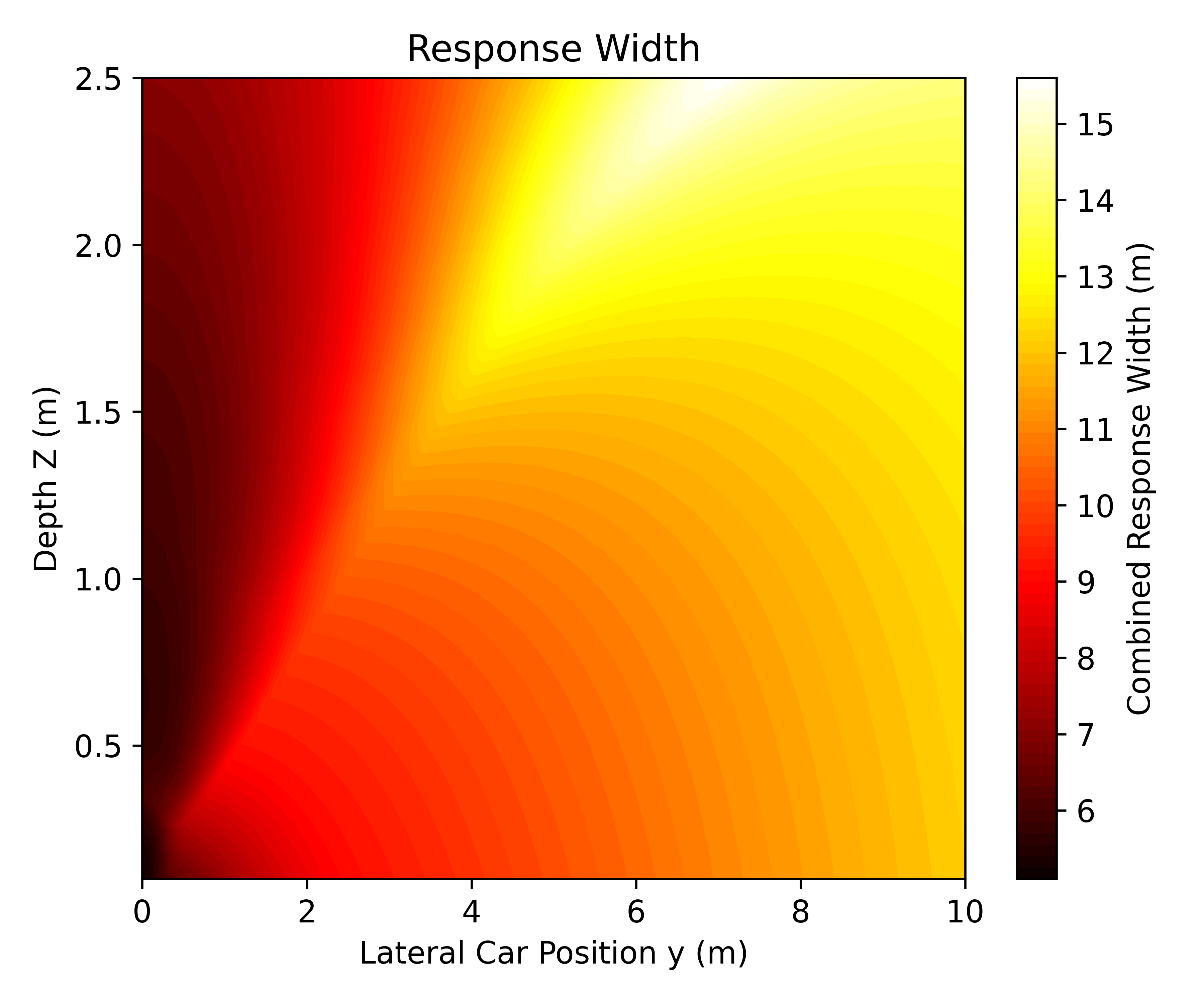}
    \includegraphics[width=1.0\linewidth]{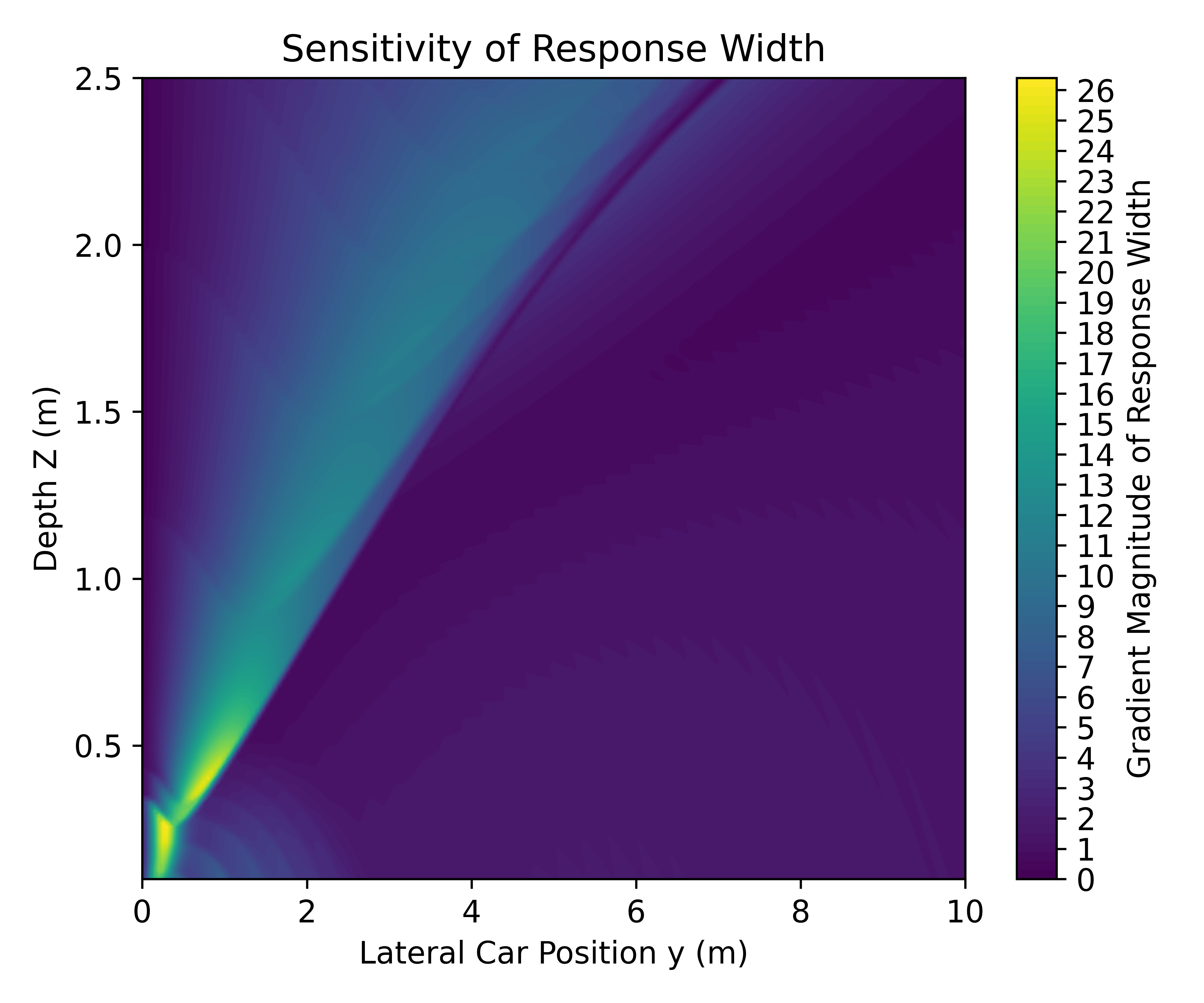}
    \caption{Top: Calculated {response} width for varying horizontal (Y) and vertical (Z) distances. The {response} width increases with both, but in a distinct pattern. Bottom: The gradient magnitude of the {response} width map, indicating sensitivity.}
    \label{fig:widthVsDistance2}
\end{figure}

The more challenging steps are estimating the fiber's burial depth and horizontal distance from the vehicle. Figure~\ref{fig:widthVsDistance} illustrates synthetic signals recorded for different geometrical conditions. 
{We define the \textit{response width} as a measure for the spatial extent, measured along the fiber-channel axis. We quantify it as the standard deviation of the absolute strain-rate profile at a fixed time.}
Fig.~\ref{fig:widthVsDistance2} shows a 2D contour in which the response width generally increases with both horizontal and vertical distance from the vehicle to the fiber, together with a sensitivity map. Around a horizontal distance of approximately $2$~m, the two opposite-sign strain-rate pulses (left negative, right positive) begin to merge into a single broader feature. It also shows that the recorded signals are highly sensitive to installation depth, which can induce significant waveform distortion with increasing installation depth. The effect of the horizontal distance is very significant for the first few meters and gradually decreases with distance. 

Our sensitivity analysis suggests that if either horizontal distance or installation depth are known, the other can be estimated from the {response} width profile. However, in the more common case where both are uncertain, aside from practical burial-depth bounds (typically between 0.5 to 1.5~m), this approach alone is insufficient for complete localization, and the full recorded signal might be required. Moreover, as Fig.~\ref{fig:widthVsDistance2} suggests, when the horizontal distance exceeds approximately three times the installation depth, the sensitivity of the {response} width to both parameters decreases dramatically. In such cases, estimating the horizontal distance becomes even more challenging, even given reasonable knowledge about installation depth. We thus conclude that the {response} width is a useful heuristic that should, nonetheless, be used carefully. 

\subsection{Correlation-Based DAS Channel Assignment}
\label{subsec:correlationMethod}
Fiber geometry estimation that directly utilizes the predicted response derived in Sec.~\ref{subsec:seismic_wavefield_fiber} 
(Eq.~\ref{eq:e_dot_das}) failed. Instead, we first conduct a coarser geometry estimation that provides a better initial model to the full optimization that uses the full properties of the signal. We note that the strain-rate pattern predicted by Eq.~\ref{eq:e_dot_das} exhibits an antisymmetric localized pulse that can be well approximated by the temporal derivative of a Ricker (Mexican-hat) wavelet, given in Eq.~\ref{eq:ricker}. This choice is motivated by its convenient analytic form and widespread use in seismic data analysis \cite{vandenende, saad2024signal, Zeroshot_denoising, DAS_up_and_down}. We therefore use the Ricker derivative not as a replacement to the full analytic model, but as a practical matched-filter template that captures the dominant antisymmetric structure. Although the exact waveform varies with load distribution (e.g., wheel spacing and vehicle length) and burial depth, empirical evaluations (Appendix~\ref{app:matchedFilter}) show that the Ricker-derivative family provides a stable and geometry-consistent initialization. This approximation is used only in the initialization stage, the second-stage optimization in Sec.~\ref{sub:synthetic_map_gen} relies on the full physics-based model.

The Ricker wavelet at scale $\sigma$ is defined as
\begin{equation}
    \psi_{\sigma}(t) = \frac{2}{\sqrt{3\sigma}\,\pi^{1/4}}
    \left(1 - \frac{t^2}{\sigma^2}\right) 
    e^{-t^2 / (2\sigma^2)}  \ ,
    \label{eq:ricker}
\end{equation}
and its temporal derivative serves as the matched-filter kernel. The scale parameter $\sigma$ directly controls the waveform spread: larger values produce broader, more slowly varying responses, whereas smaller values yield sharper, more localized features. Because neither the lateral offset nor the exact burial depth is known a priori, we treat $\sigma$ as a variable representing the {response }width broadening imposed by both installation depth and parallel offset between the fiber and the road.

For each time frame, we compute the one-dimensional convolution
between the DAS strain–rate across all channels and a bank of filters $\psi'_{\sigma_k}(-t)$ for
$\sigma_k\in[\sigma_{\min},\sigma_{\max}]$. In the presence of additive noise, matched filtering maximizes the output
signal-to-noise ratio~\cite{VanTrees1968,Proakis2007}, herein enabling reliable detection of the
crossing point. Thus, for each time frame, we select the scale $\sigma^*(t)$ and channel index $l^*(t)$ that maximize the absolute filter response. Algorithm~\ref{alg:matched_filter} summarizes this procedure, and {Fig.~\ref{fig:strain_map_range} in Section~\ref{sec:experiment} illustrates in yellow marks the resulting crossing point detections across different timeframes.}

\begin{algorithm}[t]
\caption{Matched-Filter–Based DAS Channel Assignment}
\label{alg:matched_filter}

\KwIn{
    DAS strain-rate matrix $\dot{\varepsilon}(l,t) \in \mathbb{R}^{N_\mathrm{ch} \times N_\mathrm{fr}}$;\\
    Filter scales $\{\sigma_k\}_{k=1}^{K}$;\\
    SNR threshold $T_{SNR}$
}
\KwOut{
    Peak channel index $l^*(t)$ and best scale $\sigma^*(t)$ for each time frame $t$
}

\ForEach{time frame $t$}{
    \ForEach{scale $\sigma_k$}{
        Compute matched filter response: $r_k(l,t) = (\dot{\varepsilon} * (-\psi'_{\sigma_k}))(l,t)$\;
    }
    Determine optimal scale: $k^* = \arg\max_k \max_l |r_k(l,t)|$\;
    Record peak channel: $l^*(t) = \arg\max_l |r_{k^*}(l,t)|$\;
    Record best scale: $\sigma^*(t) = \sigma_{k^*}$\;
}
Discard records with $\mathrm{SNR}(r_k(l,t)) < T_{SNR}$\;
\Return $\{l^*(t), \sigma^*(t)\}$\;

\end{algorithm}

Matched-filter detections are aggregated across frames and vehicles (see
Appendix~\ref{app:matchedFilter}) to identify a contiguous set of DAS channels that
consistently exhibit high-SNR antisymmetric signatures. This yields a \emph{channel-range estimation}, a DAS-domain mask indicating the portion of the fiber that lies within the camera’s field of view. This step is crucial when prior knowledge is missing, as it aligns real-world vehicle positions with the appropriate subset of DAS channels and eliminates irrelevant channel segments from the optimization domain.


Whereas the purpose of this step is first and foremost to estimate the X-axis of DAS channels, the matched-filter scale $\sigma^*(t)$ also provides a coarse estimate of the perpendicular distance from the vehicle to the fiber. By inverting the modeled relation between $\sigma$ and perpendicular offset (Appendix~\ref{app:matchedFilter}), we obtain an initial lateral position for each channel. Appendix~\ref{app:matchedFilter} further shows that this initialization is stable for realistic installation depth assumptions (0.6-1.1\,m). Therefore, the mapping obtained from this simplified matched filter approach can provide the initial conditions for a complete, full-physics-based optimization described next. 

\subsection{Neural Network Optimization-Based Method}
\label{sub:synthetic_map_gen}
We formulate the fiber localization as an optimization problem whose goal is to find fiber coordinates $\mathbf{S}_n$ that minimize the mismatch between the measured and the synthetic strain-rate maps generated via Eq.~\eqref{eq:e_dot_das}, i.e.:
\begin{equation}
\label{eq:total_loss}
    \min_{\mathbf{S}_n} \; \left\| \dot{\varepsilon}_{\mathrm{DAS}_{\mathbf{S}_n}}(l, t) - \dot{\varepsilon}_{\mathrm{DAS}}(l, t) \right\|^2 \ .
\end{equation}
Here, $\dot{\varepsilon}_{\mathrm{DAS}_{\mathbf{S}_n}}(l, t)$ is the synthetic map generated for the candidate fiber geometry $\mathbf{S}_n$, and $\dot{\varepsilon}_{\mathrm{DAS}}(l, t)$ is the measured DAS strain-rate map. For practical robustness, both maps are normalized to unit peak amplitude before computing the loss. This approach assumes that the vehicle trajectory, from which the synthetic maps are generated, is known - either from video footage or using a vehicle-mounted GPS. In addition, it assumes known and homogeneous subsurface parameters in the analyzed area.

In principle, one could optimize the fiber trajectory directly by adjusting the $3N$ parameters representing the coordinates of $N$ distinct fiber segments. However, this direct optimization approach was found to {{converge much more slowly, and to sometimes be unstable under noisy initialization. The corresponding ablation analysis in Appendix~\ref{app:sensitivityAnalysis} shows that, for a simple case, the MLP converges about 5 times faster than the direct optimization.}} Moreover, as our sensitivity analysis indicates, the gradients induced by changes in the vertical coordinate (depth) are substantially larger than those caused by horizontal coordinate changes, leading to difficulties in optimization. We found that these issues could be significantly mitigated by using a neural network to parameterize the fiber geometry. In this approach, the network takes as input a random latent vector $\mathbf{z} \in \mathbb{R}^{128}$ sampled from a white Gaussian distribution (a seed), and outputs a vector in $\mathbb{R}^{3N}$ representing the $x$, $y$, and $z$ coordinates of the $N$ fiber segments, which is then scaled and translated using global learnable parameters to match the proper distance units (typically, in orders of tens of meters). We employ a simple fully-connected feedforward architecture with three hidden layers, each of width $128$, and ReLU activation functions. This neural-network parameterization acts as a regularizer, constraining the solution space to smoother and more physically plausible fiber trajectories, thereby improving both stability and convergence. 

\subsection{Smoothing Regularizers}
\label{sub:smoothing_regularizers}
To ensure physically plausible and smooth reconstructions of the fiber trajectory \(\{\mathbf{S}_n\}_{n=1}^N\), we consider a few options for smoothing regularizers to the optimization. These regularizers penalize unrealistic geometric configurations, enforce gradual spatial variation, and improve numerical stability. Two regularization terms are considered, and two additional were examined and found less efficient (see Appendix \ref{app:regularizers}):

\paragraph{Angular smoothness}  
To further suppress sharp directional changes, we minimize the angular deviation between consecutive segment vectors \(\mathbf{d}_n = \mathbf{S}_{n+1} - \mathbf{S}_n\). Let \(\theta_n\) be the angle between \(\mathbf{d}_{n-1}\) and \(\mathbf{d}_n\):
\begin{gather}
    \cos\theta_n = \frac{\mathbf{d}_{n-1} \cdot \mathbf{d}_n}{\|\mathbf{d}_{n-1}\| \, \|\mathbf{d}_n\|} \ .
\end{gather}
The corresponding penalty is:
\begin{gather}
    R_{\theta} = \frac{1}{N-2} \sum_{n=2}^{N-1} \theta_n^2 \ .
\end{gather}
This term promotes smooth directional transitions without requiring explicit curvature computation.

\paragraph{Segment length uniformity}  
Given a target segment length \(\ell_0 \approx 1\), we penalize deviations from this length:
\[
R_{\ell} = \frac{1}{N-1} \sum_{n=1}^{N-1} \left( \|\mathbf{S}_{n+1} - \mathbf{S}_n\| - \ell_0 \right)^2 \ .
\]
This prevents excessive stretching or compression in the fiber representation.

\paragraph{Total regularization}  
The total regularization cost is thus:
\begin{gather} \label{eq:total_reg}
    R = \lambda_{\theta} R_{\theta} +  \lambda_{\ell} R_{\ell} \ ,
\end{gather}
where \(\lambda_{\theta}, \lambda_{\ell}\) are weighting parameters controlling the influence of each term. In our experiments, we found that \(R_{\theta}\) produced a satisfying smoothing effect and \(R_{\ell}\) contributes to the stabilization of the optimization. The values that performed best in practice were: $ \lambda_{\theta} \approx 2, \quad \lambda_{\ell} {=} 1$.

\section{Simulation Results}
\label{sec:simulations}

We simulate the motion of a car with four wheels, located at coordinates \((\pm 0.75~\text{m}, \pm 2.25~\text{m})\) relative to the vehicle's center of mass.  
The car traveled in a straight line along the $x$ axis at a constant velocity of \(v \approx 55~\text{km/h}\). The true trajectory was denoted by \(X_{\text{true}}(t_{\text{true}})\). The fiber is represented as buried underground with variable depth and distance from the straight road. The total fiber length was about \(L_{\text{tot}} = 100~\text{m}\), and it included a single vehicle pass. The DAS gauge length was \(10~\text{m}\), and the spatial sampling interval was \(1.021~\text{m}\) (matching the experimental setup in Section~\ref{sec:experiment}). The DAS temporal sampling rate was set to \(5~\text{Hz}\). We assume a Poisson ratio \(\nu = 0.25\) and shear modulus \(G = 112.5~\text{MPa}\), which are typical values for shallow, unconsolidated sediments. 

As a case study, the fiber position relative to the road follows:
\begin{gather}
    y(l) = A \cos(\omega \, l / L_{\text{tot}}) + B, \\
    z(l) = A_z \cos(\omega_z \, l / L_{\text{tot}}) + C,
\end{gather}
where \(l\) is the arc-length coordinate along the fiber.
The lateral offset \(\Delta y\) is characterized by a random undulation amplitude \(A \sim \mathcal{U}(0,\,\alpha)\),
a random frequency \(\omega \sim \mathcal{N}(0,\,1)\),
and a constant lateral shift \(B \sim \mathcal{U}(\beta/2,\,\beta)\).
The vertical offset \(z\) similarly includes a random undulation
\(A_z \sim \mathcal{U}(0,\,\gamma)\) with independent frequency
\(\omega_z \sim \mathcal{N}(0,\,1)\),
added to a burial depth \(C =0.9 \).
Here (with the $z$ axis defined as downward), \(\alpha\), \(\beta\), and \(\gamma\) are defined in meters. 

At each simulation time step, the true strain-rate response map was generated using Eq.~\ref{eq:e_dot_das}, taking into account fiber location, direction of measurement, and gauge-length effects.
{The localization accuracy, or localization error, is measured in units of length using the 3D Hausdorff distance, as presented in \ref{eq:hausd}.}

\paragraph{Noise modeling}
We add additive Gaussian background noise to recorded data:
\[
n \sim \mathcal{N}(0, \sigma_{\text{noise}}) \ .
\]
We also model errors in the visual localization system, estimating the vehicle location,
\(X_{\text{estimate}}(t_{\text{estimate}})\), with Gaussian noise in both position:
\[
X_{\text{estimate}} \sim \mathcal{N}(\mu_X, \sigma_X) \ .
\]
In practice, typical temporal sampling errors are on the order of milliseconds and are therefore negligible, so we assume $t_{\text{estimate}}$ is deterministic. To isolate these effects, we set $\sigma_X=0$, except in the localization analysis section.

\paragraph{Anchors and initialization}
The number of exact spatial \textit{anchors}, denoted \(A\), was selected between 1 and 5, and anchor locations were sampled uniformly along the fiber. For simplicity, the pre-optimization fiber location was set via 3D cubic-spline interpolation of the anchor set. In the single-anchor case, points were initialized randomly around that anchor. 
{In most typical cases, anchor locations have some associated uncertainty. For example, a practical scenario could be estimating manhole locations with limited accuracy. Therefore, we distinguish between two cases: a \textit{hard anchor}, which imposes a constraint on the fiber edge location, and a \textit{soft anchor}, which only determines the initialization for the optimization. It is important to note that in the hard anchor case, for any given error $\varepsilon$ in the anchor location, the final fiber localization error (Eq. \ref{eq:hausd}) is at least $\varepsilon$. However, this is not the case for the soft anchor, where the optimization scheme can correct the initial anchor location error. In Appendix \ref{app:anchorsUncertainty}, we present a comparison between the two cases.} As a baseline, two hard anchors were used in the experiments.

\paragraph{Optimization procedure}
The fiber reconstruction was optimized using the neural-network framework in Section~\ref{sub:synthetic_map_gen}, with Adam, a learning rate of \(0.005\), and \(20\) optimization steps. Because depth (the \(z\) coordinate) exhibits stronger sensitivity, we used anisotropic learning rates by attenuating \(z\)-gradients by two orders of magnitude. In addition, updates of the \(z\) coordinate were enabled only during the second half of optimization. The objective was Eq.~\ref{eq:total_loss} with regularization from Eq.~\ref{eq:total_reg}. In these simulations, we used only segment-length regularization \(\lambda_{\ell}=1.0\) and \(\lambda_{\theta} \approx 0\).

\subsection{Parameter Study}
\label{subsec:parameter_study_sim}

\begin{figure}[h]
\centering
\includegraphics[width=1.0\linewidth]{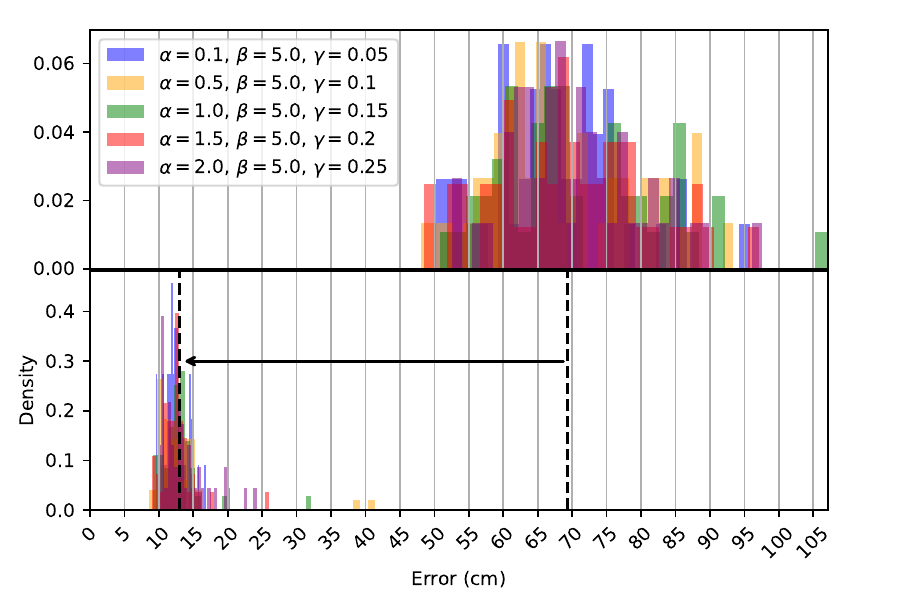}
\caption{Localization accuracy under different initial fiber conditions. Top: before optimization (after interpolation-based initialization). Bottom: after optimization. Results are aggregated over 50 trials, the average improvement is almost $60$ cm in Hausdorff distance.}
\label{fig:differentXYhist}
\end{figure}

We examined the following factors influencing the convergence of the optimization algorithm using simulated data. In all { the following use }cases, we set $(\alpha, \beta, \gamma) = (1, 5, 0.1)$, used two anchors, assumed negligible noise in both localization and background measurements, and used the correct Poisson ratio. {Figure \ref{fig:differentXYhist} shows the histograms of the optimization results for different values of $(\alpha, \beta, \gamma)$.}

\begin{figure}[h]
\centering
\includegraphics[width=1.0\linewidth]{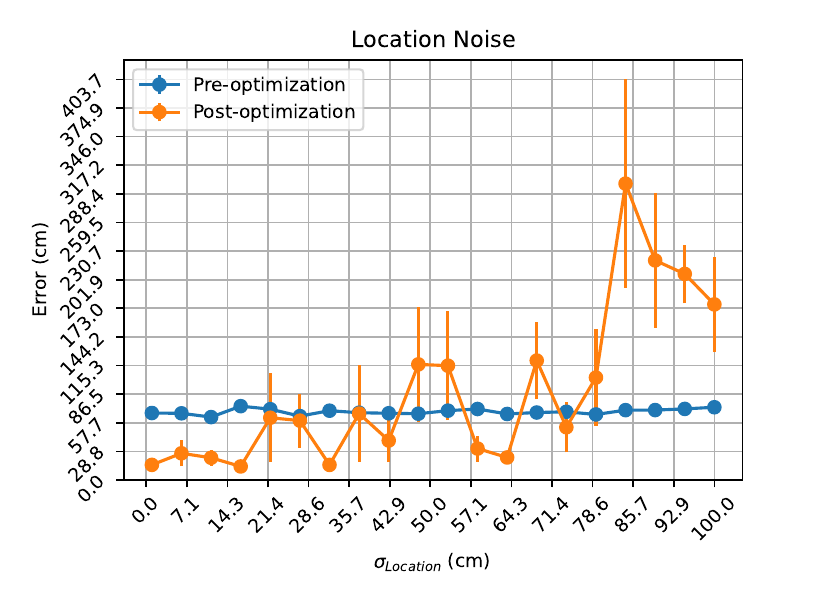}
\caption{Localization error as a function of spatial measurement noise. 
The results show a clear tolerance threshold: performance degrades rapidly beyond $\approx 35$ cm. Values are reported as mean $\pm$ standard deviation over 10 trials. The fiber was buried at a depth of $90$ cm, with an average horizontal offset of $3.75$ m, and two anchors were used.}
\label{fig:heatmap_loc_noise}
\end{figure}

\paragraph{Spatial measurement accuracy.}
We simulated multiple levels of spatial uncertainty in vehicle trajectory estimation, representing realistic errors from vehicle-mounted GPS or camera tracking. These errors directly propagate into the calibration loss. Position noise was modeled as Gaussian perturbations. {Figure \ref{fig:heatmap_loc_noise} shows that a}s expected, larger measurement noise increased interpolation error and degraded localization accuracy. Under constant vehicle speed, temporal uncertainty (sampling-time error) maps linearly to spatial uncertainty.

\begin{figure}[h]
\centering
\includegraphics[width=1.0\linewidth]{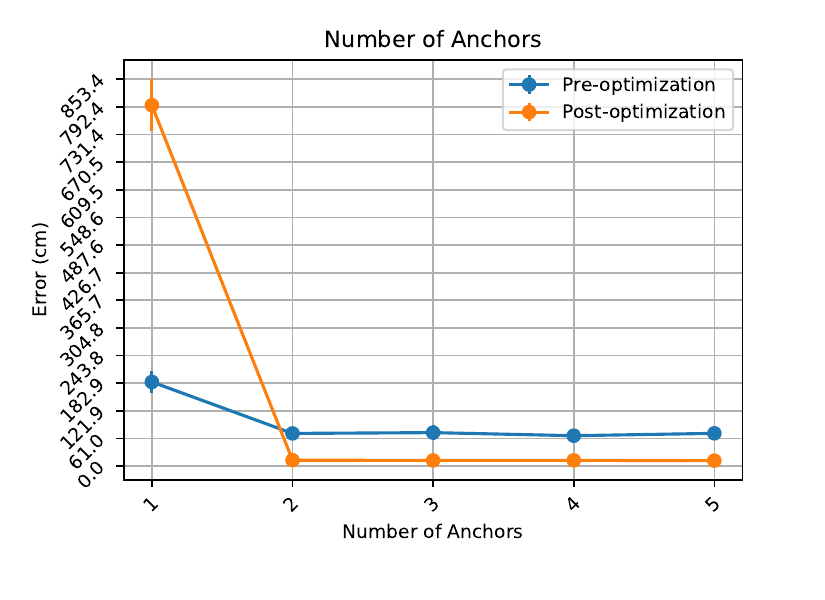}
\caption{Effect of the number of anchors on localization error. Results are reported as mean $\pm$ standard deviation over 10 trials. Accuracy improves substantially once two or more anchors are available. The fiber was buried at a depth of $90$ cm, with an average horizontal offset of $3.75$ m.}
\label{fig:heatmap_anchors}
\end{figure}

\paragraph{Number and placement of anchor points.}
In the optimization framework (Sec.~\ref{sec:method}),  
we optionally incorporate a sparse set of known fiber locations (anchors) to constrain the reconstruction {(Hard anchors)}. { As seen in Figure \ref{fig:heatmap_anchors},  w}e varied the number of anchors and their spatial distribution along the fiber to evaluate effects on convergence speed and reconstruction fidelity. Anchor locations were sampled uniformly at random along fiber length. Results show that at least two anchor points along the fiber are required to achieve reliable performance for a 100-m long fiber.

\paragraph{True fiber installation depth and lateral offset}
Here, we analyze multiple experiments, each with a different true fiber location, and evaluate the difference between the pre-optimization (initialization based on the anchors), and post-optimization distances from the estimated fiber to the true fiber. Depth errors exhibit a stronger nonlinear effect on reconstruction quality than horizontal offsets, consistent with the gradient-based sensitivity results in Appendix~\ref{app:sensitivityAnalysis}. This analysis assumes initialization within a reasonable pre-optimization range (sub-meter scale).

\begin{figure}[h]
\centering
\includegraphics[width=1.0\linewidth]{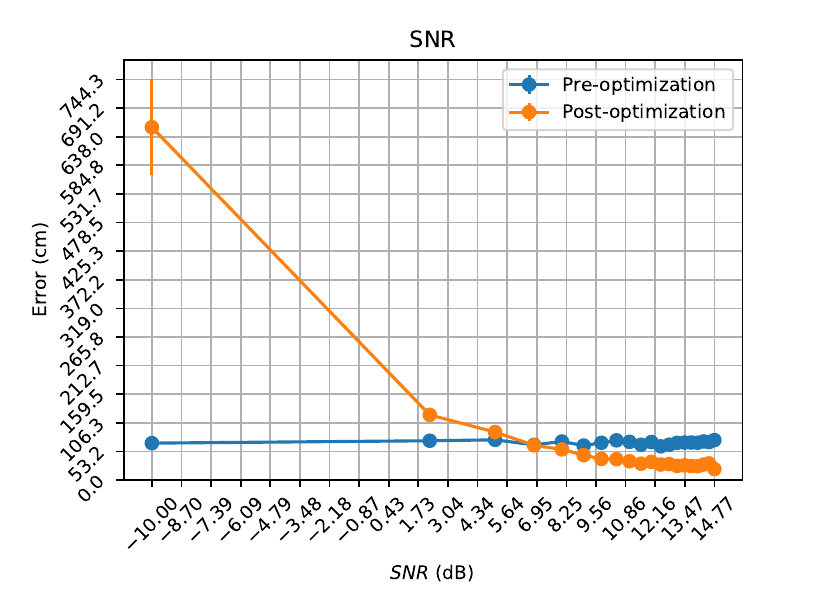}
\caption{Localization accuracy as a function of SNR in the DAS strain-rate map. Performance remains stable down to approximately 7 dB, below which accuracy degrades. Results are reported as mean $\pm$ standard deviation over 10 trials. The fiber was buried at a depth of $90$ cm, with an average horizontal offset of $3.75$ m, and two anchors were used.}
\label{fig:heatmap_snr}
\end{figure}

\paragraph{Signal-to-noise ratio (SNR)}
The DAS strain-rate maps are affected by background noise from environmental and anthropogenic sources, instrumentation, and modeling uncertainty. { Figure \ref{fig:heatmap_snr} shows a systematic variation of the} noise amplitude (white-noise assumption) to simulate different SNR conditions. Higher noise reduces the visibility of strain-rate patterns and increases ambiguity in trajectory recovery. Notably, moderate noise levels (above \(\sim 7\) dB SNR) did not significantly impair reconstruction, indicating resilience of the neural-network-based optimization.

\begin{figure}[h]
\centering
\includegraphics[width=1.0\linewidth]{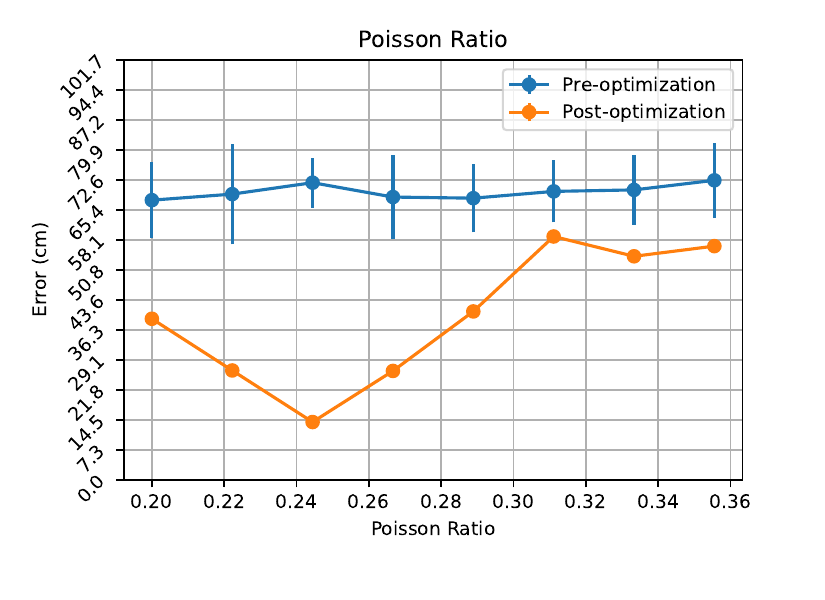}
\caption{Localization accuracy as a function of Poisson ratio $\nu$ in the DAS strain-rate map. The true underground has $\nu=0.25$, but the optimization was conducted using different assumed values to model the lack of subsurface information. Results are reported as mean $\pm$ standard deviation over 10 trials. The fiber was buried at a depth of $90$ cm, with an average horizontal offset of $5$ m, and two anchors were used.}
\label{fig:poisson}
\end{figure}

\paragraph{Poisson Ratio}
In our method, the absolute value of the shear modulus has, as long as it is homogeneous, no influence because the strain map is normalized. However, the value of the Poisson ratio does influence the shape of the recorded signal. We therefore examine the effect of assuming an incorrect Poisson ratio during optimization. The results are shown in Fig.~\ref{fig:poisson}. As expected, the best performance is obtained when the correct Poisson ratio is used, and deviations from that value degrade performance. As a rule of thumb, an error of $\Delta \nu \approx 0.01$ causes an error of about $5$ cm. Typical Poisson values are in the range of 0.2-0.4.

\section{Experimental Results}
\label{sec:experiment}

We conducted two experiments on Klausner Street near Tel Aviv University, Israel.
The first experiment is a previously published dataset \cite{dataset} acquired during a week-long campaign in March 2023 and includes synchronized DAS and camera recordings. Object detection and tracking in video were performed using YOLOv11 (Ultralytics) \cite{redmon2016you, yolo11_ultralytics} (see Appendix \ref{app:YOLOanalysis} for model analysis). The second experiment took place in August 2025 and consisted of a 1-hour recording synchronized with 5 Hz GPS measurements from a 4.4-ton vehicle driving along the same street (see Appendix~\ref{app:gpsTraj} for the measured GPS trajectory). GPS precision varied between \(\sigma_{\text{GPS}}=4-7\) cm.

DAS measurements were continuously recorded using a Silixa iDAS V2 IU, with 1.021 m channel spacing and a fixed 10 m gauge length. The fiber follows the street for approximately 250 m. Two fiber anchors within the camera coverage area were manually located using prior calibration sledgehammer blows targeting manholes where available. 

\begin{figure}[t]
    \centering
    \includegraphics[width=1.0\linewidth]{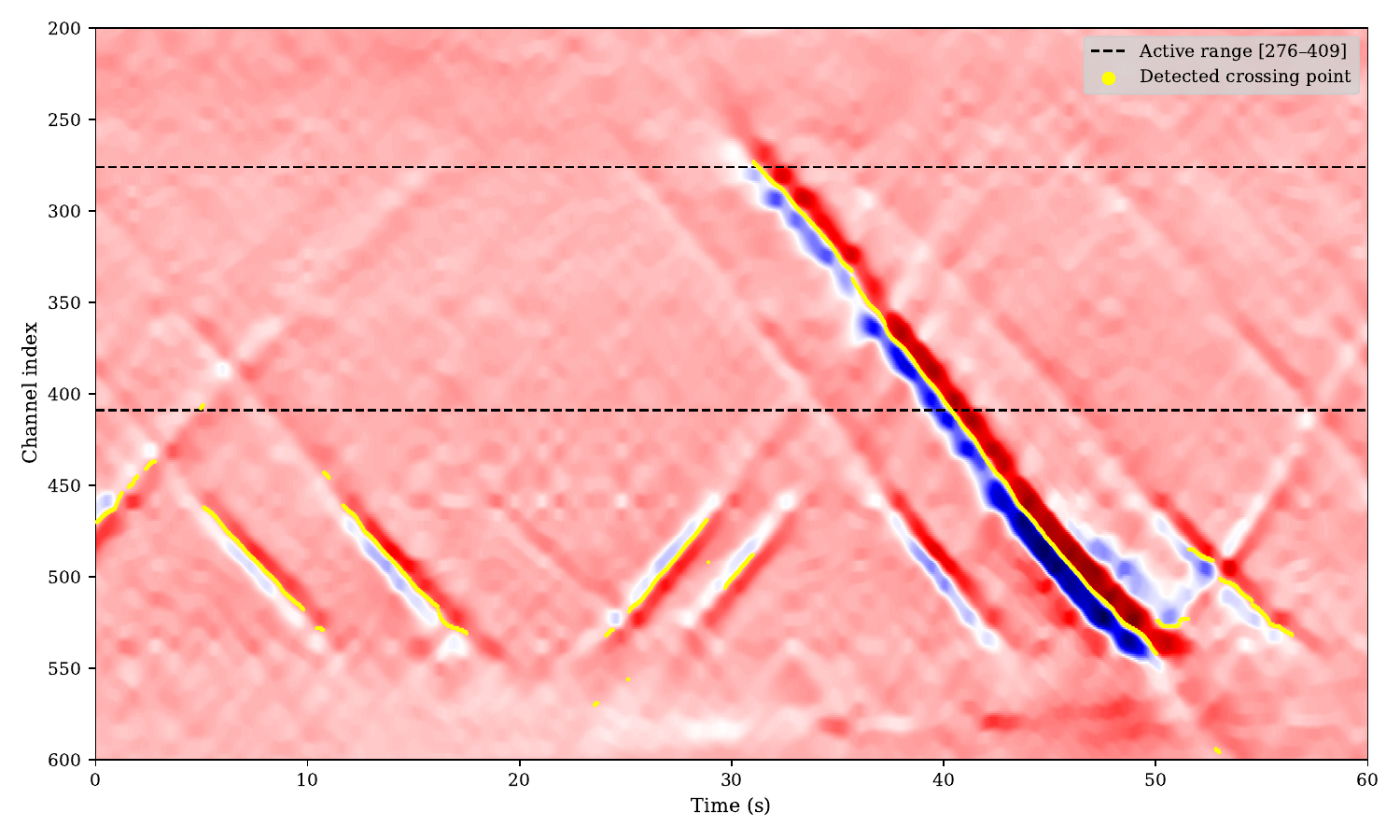}
    \caption{Strain-rate map over the full DAS data, including channels that are not near the road. The dashed vertical lines mark the estimated active range $[276,\,409]$ that overlaps with the visible road segment. This region exhibits coherent vehicle-induced signatures, confirming the calibration result. {{Yellow markers indicate the per-frame matched-filter crossing-point detections.}}}
    \label{fig:strain_map_range}
\end{figure}

\textbf{Step 1 - Matched filtering} Before performing fiber localization, it is necessary to determine which subset of DAS channels corresponds to the visible portion of the road captured by the camera. This calibration step aligns the DAS and vision coordinate systems and ensures that subsequent processing uses only physically relevant data. The estimated \emph{active channel range} of the fiber within the visible scene was found to be $[276,\,409]$. This closely matches the manually calibrated range of $[280,\,410]$, demonstrating strong agreement between the automatic matched-filter procedure and the ground-truth calibration (Fig.~\ref{fig:strain_map_range}).

With the active channel range $[\ell_{\min},\ell_{\max}]$ established, we further exploit the matched-filter output to obtain an \emph{initial guess} for each channel’s lateral offset. Specifically, for every assigned frame within $[\ell_{\min},\ell_{\max}]$, we invert the coupling curve $\sigma(y;z)$ (Sec.~\ref{subsec:seismic_wavefield_fiber}) to map the observed filter scale $\sigma_\ell$ to a radial distance $\rho_\ell=f^{-1}(\sigma_\ell; z)$. Assuming a constant burial depth $z$, these $\rho_\ell$ values yield per-channel circular loci $\mathbf{p}_\ell({\phi_\ell})$ as in \eqref{eq:circular_param}, which seed the trajectory reconstruction (Sec.~\ref{sec:method}).

\textbf{Step 2 - Neural-network optimization} Having established the channel range and depth-dependent initialization, we next generate the synthetic strain-rate map and optimize the fiber trajectory within this calibrated region.

\begin{figure}
    \centering
    \includegraphics[width=1.0\linewidth]{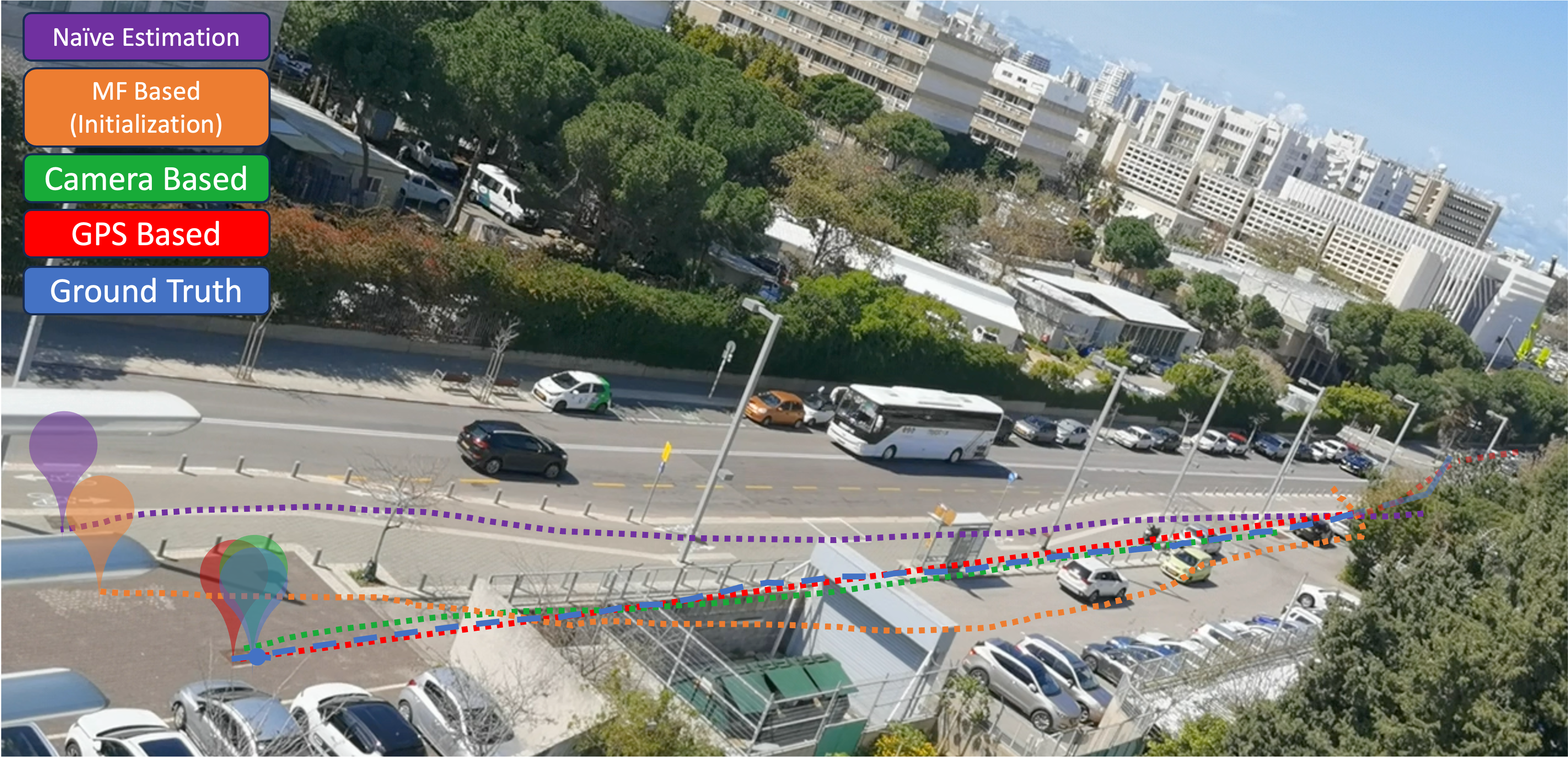}
    \caption{Camera view of the street covering approximately 70 m. Naive estimation (purple) indicates the initial fiber guess. GPS based (red) and Camera based (green) curves show the reconstructed fiber trajectories. Red and green markers denote anchor points used during optimization. MF based initialization (orange) denotes correlation based initialization points, assuming a fixed depth of $z=0.9\,\mathrm{m}$. Ground truth (blue) represents the high-resolution utility mapping provided by Tel Aviv University.}
    \label{fig:ExperimentStreet}
\end{figure}

Both camera- and GPS-based experiments were conducted using a single heavy vehicle pass along the fiber. Although averaging across multiple passes may further improve robustness, we found that a single pass is sufficient when SNR is high.
Processing and evaluation were performed on a single machine. The resulting fiber localization is shown in Fig.~\ref{fig:ExperimentStreet}, together with a comparison to high-resolution utility mapping provided by Tel Aviv University {which was supplied by a contractor using electromagnetic-based mapping infrastructure techniques}. Both estimates lie within a Hausdorff distance of approximately 35–50 cm from the existing utility mapping, while differing from each other by about 20 cm {(see Appendix \ref{app:expErrorEval} for the error profile along the fiber)}. The GPS estimation also shows the fiber crossing the road, which is also validated by utility maps beyond the field of view. However, in this area video coverage was poor so we did not consider this section for error analysis. 

\section{Discussion}

Our proposed methodology shows that a physics-based optimization approach using quasi-static seismic signals generated by known vehicles can be used to accurately reconstruct fiber-optic routes in proximity (several meters) to roads. Ground-truth information about vehicle trajectory can be obtained either by video processing or a vehicle-mounted GPS measurement, but our formulation is agnostic to the origin of the information and can use other measurements as well. 

Mapping the fiber trajectory along the direction of the road is relatively easy, as has also been shown by previous studies. However, the inherent coupling between horizontal offset and burial depth makes separate estimation of these quantities challenging without a priori information. Therefore, we opt for a full physical modeling of the recorded signal, and use it for gradient-based fiber trajectory optimization. This, in turn, requires reasonable initial conditions, which can be obtained by an approximated matched-filter approach that only analyzes first-order statistics of the vehicle-induced DAS patterns. Despite its simplicity, it has the benefit of being a broad-search optimization approach that is less sensitive to local minima than gradient-based optimization. Overall, our analysis shows that under reasonable acquisition conditions (DAS SNR of above 7 dB and standard deviation of less than {$35$~cm} in the vehicle trajectory estimation), the optimization process converges to errors in the range of tens of centimeters, which is sufficient to avoid damage to infrastructure during excavation.  

In general, analyzing heavier vehicles such as buses or trucks may be beneficial as the generated quasi-static signals are, to first order, proportional to the vehicle weight. The SNR is thus expected to be higher and, similarly, performing the optimization for multiple vehicles and analyzing the ensemble results may improve the stability of the estimations. We note that our approach used a point load approximation, which may not be justified for {heavier vehicles, or more than 4 wheels, as we show systematically in Appendix \ref{app:PLSapprox}}. The field data examples converged successfully using a 4.4-ton van. Our analysis shows that for a 10-m gauge length and 4.5-m axle distance, the point load approximation is justified. In any case, accounting for the full load distribution, if justified given the axle length and gauge length of the DAS interrogator, is straightforward to implement given knowledge of the vehicle type from video inputs. 

A more limiting set of assumptions relates to subsurface homogeneity and knowledge of its properties. The absolute value of the shear modulus $G$ is not necessary as data are normalized to account for coupling and choice of optical parameters. However, the Poisson ratio $\nu$ directly influences the shape of recorded data. Synthetic tests indicate that a 0.01 error in the Poisson ratio leads to approximately 5 cm error in the fiber localization. For both parameters, unmapped lateral variations will skew recorded data in a way that cannot be accounted for by the physical model. We thus first recommend performing optimization for long fibers in an iterative fashion, processing relatively short segments (on the order of 100 m) individually, thus handling data from areas with similar lithology. In an iterative approach, information from the end of the previous segment informs the start of the next segment to optimize. In addition, estimation of subsurface properties from ambient noise DAS is a well-studied topic with established workflows. These can be used to strongly constrain the Poisson ratio as well as help detect areas of sharp lithological changes. In these cases, both sides of the boundary should be processed separately. { In Appendix \ref{app:FiberScalability} we show an example where we localize 1 km fiber using this approach.}

From an optimization perspective, the main limitation of our proposed technique is that different scenarios still require moderate hyperparameter tuning to achieve optimal convergence. Among the tested regularizer, \textit{segment-length uniformity} has the strongest impact, followed by \textit{angular smoothness}. Optimization in depth remains the most challenging component, as shown in Fig.~\ref{fig:differentXYhist}. Generally speaking, utility information about installation depth can be, if reliable, highly valuable and significantly improve convergence. As a gradient-based approach, the optimization process is sensitive to initial conditions and prior information. In our experiments, constraining at least two fiber locations accurately was typically sufficient to ensure convergence for a 100-m long fiber, even for non-straight geometries (Fig.~\ref{fig:heatmap_anchors}). In practice, this is most likely achieved by detecting manholes or optical cabinets, whose position can be measured directly. In urban environments, assuming typical distances of 100 m between such elements is realistic, and we can thus expect convergence. We also assume that spool sections are not present in the optimization interval, as they do not adhere to the model. In practice, this is usually not restrictive, because spools are easy to detect using ambient seismic data \cite{bukharin2023ambient}. Detected spool regions can then be excluded from the optimization or treated separately. We note that although we experimented with several other types of loss functions, we determined that L2 was sufficient. Nevertheless, there remains potential for better loss functions, which we leave to future studies. These may improve convergence and reduce the dependency on initial conditions and a priori knowledge. 

{Compared to other related studies, our method shows the first example of sub-meter accuracy. This resolution is obtained from using the full spatio-temporal DAS response instead of the peak channel. A few previously mentioned studies have demonstrated accuracies of up to one meter using different approaches, such as fiber localization with a seismic source (a hammer strike) by Liu et al.\cite{liu2020new}, but they are neither passive nor scalable. Among vehicle-driven methodologies, Bukharin et al. \cite{bukharin2023ambient} provide only an indirect estimation of the fiber path, as they actually detect manholes and then linearly interpolate between them, achieving an accuracy of 75\%–90\%. Biondi et al. \cite{biondi2023geolocalization} use a similar vehicle–GPS synchronization technique, yielding a 1-to-5-meter longitudinal (along-track) error, but suffering from severely limited transverse (cross-road) resolution.}
{{Additional quantitative comparisons with alternative methods are provided in Appendix~\ref{app:MethodsComparison}.}}

It is important to note that a relatively simple road geometry, with a single lane in each driving direction, was studied. In addition, we only cover a relatively short segment of about 100 m { which, in our experiment, was covered with both GPS and synchronized camera}. Even under these conditions, a relatively complex and parameter dependent optimization workflow was required. Therefore, { even though our initial results are encouraging,} further studies are needed to validate the generality {and the robustness} of this approach and how it can be applied over longer road segments. Such studies will also be important in testing the assumptions about subsurface homogeneity and the prior knowledge of its properties.

\section*{Data availability}
The GPS dataset is available at Zenodo \url{https://doi.org/10.5281/zenodo.19497741}.
The camera-based dataset used in the first experiment is available at \url{https://doi.org/10.5281/zenodo.15869300}.

\section*{Acknowledgments}
This research was supported by the Israeli Ministry of Innovation, Science, and Technology, under grant no 1001953425 (Smart Transportation and Infrastructure), the Blavatnik Artificial Intelligence and Data Science Fund, and the Shlomo Shmeltzer Institute for Smart Transportation at Tel Aviv University. K.C. is supported by the VATAT (PBC) Fellowship for Outstanding PhD Students in Data Science. The authors thank Exodigo Ltd., and in particular Pavel Sinitsyn, for their help in the experiment. We also thank Omer Shamir for help with DAS data acquisition.

\bibliographystyle{IEEEtran}
\bibliography{references}

\newpage

\appendices






\section{Fiber System of Coordinates}
\label{app:FiberSystem}

In order to analyze seismic-induced deformation from the fiber’s point of view, we define a local orthonormal frame $\left\{\hat{e}_x(l), \hat{e}_y(l), \hat{e}_z(l)\right\}$ at each point along the fiber using the Frenet-Serret construction:
\begin{align}
    \hat{e}_x(l) &= \frac{d\vec{S}/dl}{\left\|d\vec{S}/dl\right\|} && \text{(Tangent vector)} \label{eq:tangent}  \ ,\\
    \hat{e}_y(l) &= \frac{d^2\vec{S}/dl^2 - \left( \frac{d^2\vec{S}}{dl^2} \cdot \hat{e}_x(l) \right)\hat{e}_x(l)}{\left\|d^2\vec{S}/dl^2 - \left( \frac{d^2\vec{S}}{dl^2} \cdot \hat{e}_x(l) \right)\hat{e}_x(l) \right\|} && \text{(Normal vector)} \label{eq:normal}   \ , \\
    \hat{e}_z(l) &= \hat{e}_x(l) \times \hat{e}_y(l) && \text{(Binormal vector)}   \ .\label{eq:binormal}
\end{align}

Together, this local frame forms a smoothly varying orthonormal coordinate system attached to the fiber, suitable for projecting external fields (e.g., displacements or forces) onto the fiber axis. This transformation is especially important when the fiber path is curved or arbitrarily shaped, as it generalizes strain modeling to arbitrary fiber geometries.

\section{Matched Filter Algorithm and Applications}
\label{app:matchedFilter}

This appendix provides implementation details and practical extensions of the correlation-based DAS channel assignment method introduced in Sec.~\ref{subsec:correlationMethod}.

\subsection{Matched-Filter Implementation}

For each video frame in which a vehicle is detected, the corresponding DAS segment is convolved with a bank of derivative-Ricker filters $\psi'_{\sigma}(-t)$, parameterized by $\sigma \in [\sigma_{\min}, \sigma_{\max}]$. In our experiments, we employ a filter bank spanning $\sigma \in [2, 20]$ with a stride of $0.5$.
The filter yielding the maximal peak response determines the assigned channel $\ell_t$ and its associated scale $\sigma_t$.
These quantities respectively indicate the fiber channel nearest to the vehicle and the effective response width.

\begin{figure}[t]
\centering
\includegraphics[width=0.9\linewidth]{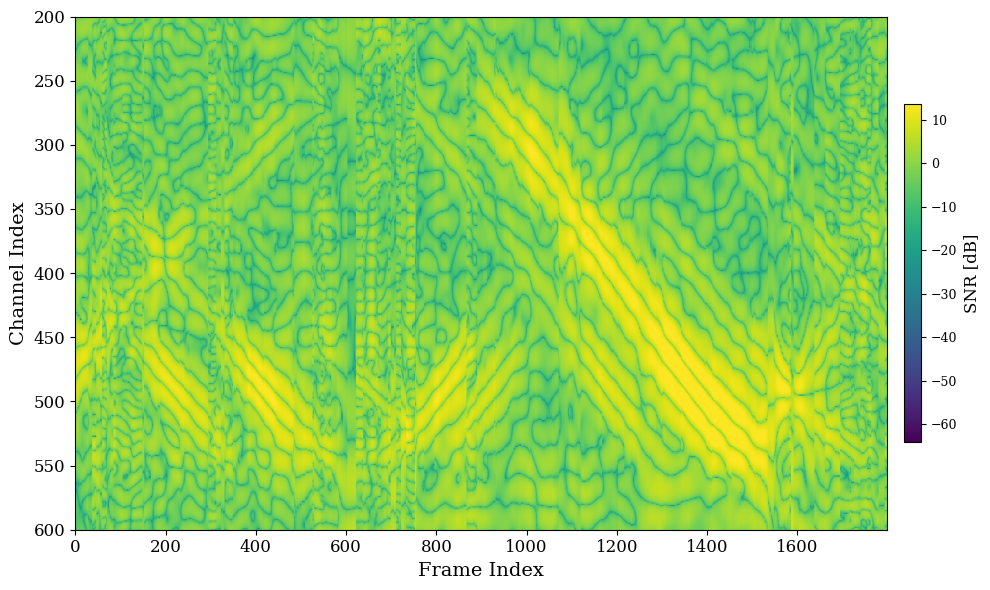}
\caption{SNR map obtained after matched filtering and MAD-based normalization.
Bright regions correspond to high-SNR vehicle responses, with incoherent background energy strongly attenuated.}
\label{fig:snr_map_example}
\end{figure}

\begin{figure}[t]
\centering
\includegraphics[width=0.9\linewidth]{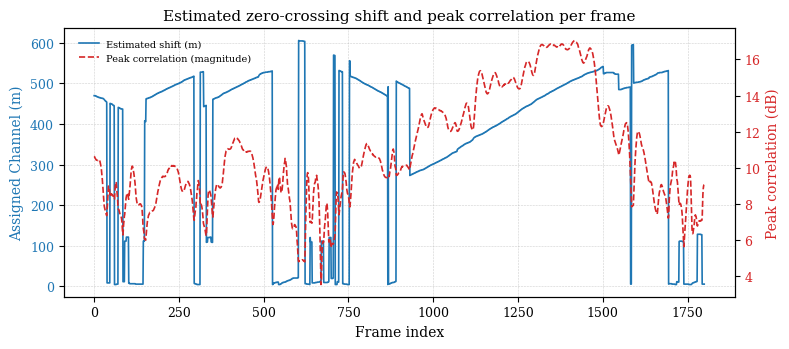}
\caption{Example of per-frame channel assignments corresponding to Fig.~\ref{fig:snr_map_example}.
Linear segments mark the temporal trajectory of a detected vehicle across the fiber optic channels.}
\label{fig:das_assignment_trace}
\end{figure}

The matched-filter outputs $r(\ell,t)$ are normalized to form signal-to-noise ratio (SNR) maps:
\begin{equation}
\mathrm{SNR}(\ell,t)
= 10 \log_{10}\left(\frac{|r(\ell,t)|}{\hat{\sigma}}\right)\,[\mathrm{dB}],
\label{eq:snr_map}
\end{equation}
where $\hat{\sigma}$ denotes a robust noise estimate computed via the median absolute deviation (MAD)~\cite{huber1981robust}.
This normalization suppresses non-stationary background fluctuations and highlights coherent, vehicle-induced strain features.
Figures~\ref{fig:snr_map_example} and~\ref{fig:das_assignment_trace} show representative SNR distributions and the resulting channel assignments. Based on the analysis in Fig.~\ref{fig:heatmap_snr}, the SNR threshold used in our experiments was set to 7 dB.

\subsection{Noise Suppression and Region Cleanup}

Prior to matched filtering, a 2-D FK filter is applied to maintain signals with phase velocities between $9-90$ km/h and frequencies below $1$ Hz.  This is followed by a 2-D median filter applied to remove impulsive artifacts.

\subsection{Directional Consistency and Global Refinement}

On the video side, object tracks are obtained using a YOLO-based detector and mapped to real-world $(x,y)$ coordinates through homography calibration.
For each frame, the assigned DAS channel $\ell_t$ and scale $\sigma_t$ are associated with the corresponding tracked vehicle.
Directional consistency is then enforced: along a given trajectory, channel indices must vary monotonically—decreasing for leftward motion and increasing for rightward motion—thus eliminating occasional sign inversions caused by overlapping sources or occlusions.

\begin{figure}[t]
\centering
\includegraphics[width=\linewidth]{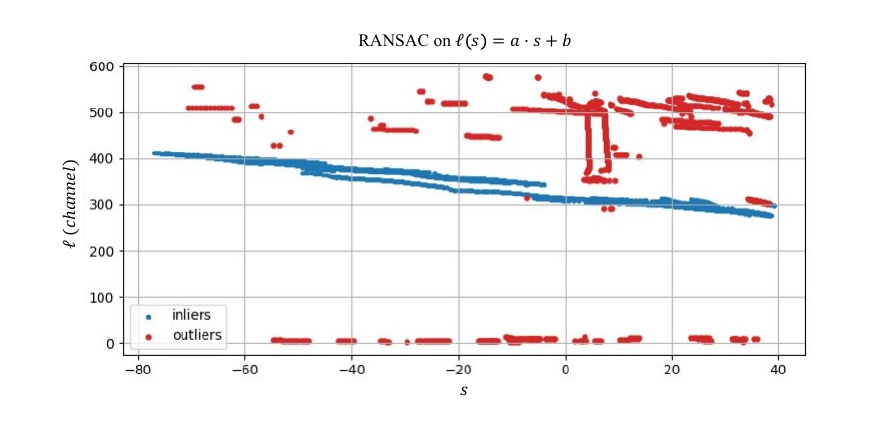}
\caption{RANSAC regression between the projected vehicle coordinate $s$ and the assigned DAS channel.
Blue markers denote inliers forming a coherent trend corresponding to the visible fiber segment,
red markers indicate outliers rejected by the model.}
\label{fig:ransac_s}
\end{figure}

After aggregating assignments across the entire sequence, a robust global refinement is applied using RANSAC~\cite{fischler1981ransac}.
Vehicle positions $(x,y)$ are projected onto their first principal component coordinate $s$, and an approximate linear relation
\begin{equation}
\ell(s) \approx a s + b
\label{eq:ransac_model}
\end{equation}
is fitted between projected position and channel index.
Outliers inconsistent with the dominant linear trend are rejected, yielding a coherent inlier set that delineates the active portion of the fiber observed by the camera.
The resulting channel range $[\ell_{\min}, \ell_{\max}]$ defines the region of interest for the subsequent fiber-geometry optimization (Fig.~\ref{fig:ransac_s}).

\subsection{Initialization via Coupling-Derived Offsets}

\begin{figure}[t]
\centering
\includegraphics[width=0.9\linewidth]{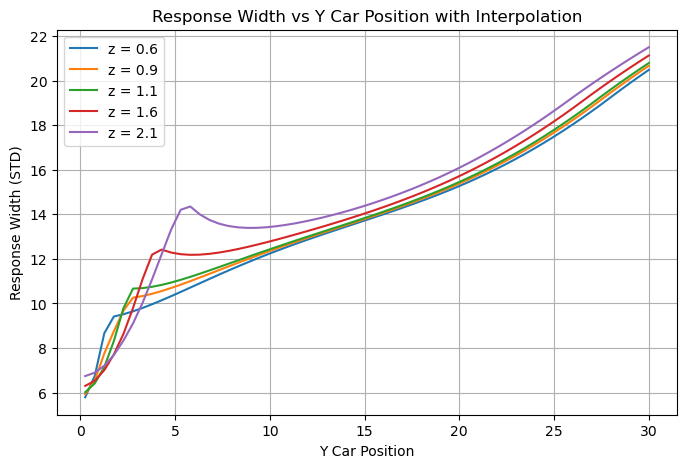}
\caption{Modeled dependence of the matched-filter width $\sigma$ on lateral offset $Y$ for several fixed depths $Z$. 
Shallower fibers exhibit a monotonic coupling decay, while deeper configurations develop secondary lobes.}
\label{fig:coupling_vs_y}
\end{figure}

The RANSAC-based regression in Fig.~\ref{fig:ransac_s} provides a reliable estimate of the active channel range observed by the camera. 
To initialize the subsequent trajectory reconstruction, we exploit the analytical coupling behavior between the strain amplitude and the lateral offset of the surface load relative to the buried fiber. 
Assuming a fixed burial depth $Z$ and homogeneous elastic parameters, the matched-filter width $\sigma$ serves as a surrogate for the load–fiber coupling strength. 
Figure~\ref{fig:coupling_vs_y} depicts the modeled dependence of $\sigma$ on lateral distance $Y$ for representative depths $Z \in \{0.6,1.1,1.6,2.1\}\,\mathrm{m}$, as derived from the analytical strain formulation in Sec.~\ref{sub:sensAnalysis}.

For each channel within the calibrated range $[\ell_{\min}, \ell_{\max}]$, the observed $\sigma_\ell$ is mapped to a corresponding radial distance 
\begin{equation}
\rho_\ell = f^{-1}(\sigma_\ell;Z),
\end{equation}
where $f(\cdot)$ represents the modeled $\sigma(Y;Z)$ relation.
At shallow and moderate depths ($Z \leq 1.1$~m), the coupling function $f$ is strictly monotonic increasing, yielding a unique and stable inversion of the observed matched-filter width $\sigma_\ell$ into lateral offset $\rho_\ell$. However, as depth increases ($Z \geq 1.6$~m), $f$ becomes locally non-bijective, introducing ambiguity into the inversion process. Specifically, multiple lateral positions may produce similar coupling responses, making the mapping $\sigma_\ell \mapsto \rho_\ell$ inherently ill-posed in this regime.

To preserve determinism and stability during initialization, we constrain the inversion to the principal monotonic segment of $f$—corresponding to the near-field response—thereby enforcing a single-valued mapping:
\begin{equation}
\rho_\ell = 
\underset{Y \in [0,Y_{\max}]}{\operatorname{argmin}}\,
|\sigma(Y;Z)-\sigma_\ell|.
\label{eq:monotonic_rho}
\end{equation}
This choice avoids spurious solutions but may lead to increasing deviation from the true fiber geometry at greater depths. We view this tradeoff as a principled limitation of the current model, and addressing the full inversion in the presence of non-bijective coupling remains an open challenge for future work.

The resulting offsets $\rho_\ell$ define local circular loci for each channel:
\begin{equation}
\mathbf{p}_\ell(\phi_\ell) = (x_\ell, y_\ell) + \rho_\ell (\cos\phi_\ell, \sin\phi_\ell),
\label{eq:circular_param}
\end{equation}
where $\phi_\ell \in [0, 2\pi)$ is an angular parameter describing the unknown orientation of the fiber in the local plane. 
These loci serve as initial conditions for reconstructing a continuous and spatially coherent 2D trajectory.

To obtain a smooth curve consistent with the regularization terms introduced in Sec.~\ref{sub:smoothing_regularizers}, we minimize a reduced form of the total regularization cost:
\begin{equation}
\mathcal{L}(\boldsymbol{\phi}) = 
\lambda_{\kappa} R_{\kappa}^{(2D)} 
+ \lambda_{\theta} R_{\theta}
+ \lambda_{\ell} R_{\ell},
\label{eq:init_cost}
\end{equation}
where $R_{\kappa}^{(2D)}$ {(Eq.~\eqref{eq:kappa})} penalizes excessive curvature (torsion $\tau_n$ omitted in 2D), $R_{\theta}$ suppresses sharp directional changes between successive segments, and $R_{\ell}$ enforces uniform segment lengths along the reconstructed path.
The channel-wise orientations $\boldsymbol{\phi}$ are optimized using the L-BFGS-B algorithm \cite{byrd1995lbfgsb} under the condition of $\boldsymbol{\phi}\in[0,2\pi)$.
The resulting coordinates are then
\begin{equation}
(x_\ell^\ast, y_\ell^\ast) = (x_\ell + \rho_\ell \cos\phi_\ell^\ast,\,
y_\ell + \rho_\ell \sin\phi_\ell^\ast),
\end{equation}
forming a smooth 2D trajectory estimate of the buried fiber. 
A low-order polynomial is finally fitted to $\{(x_\ell^\ast, y_\ell^\ast)\}$ to enforce global consistency and suppress residual drift.

\begin{figure}[t]
\centering
\includegraphics[width=\linewidth]{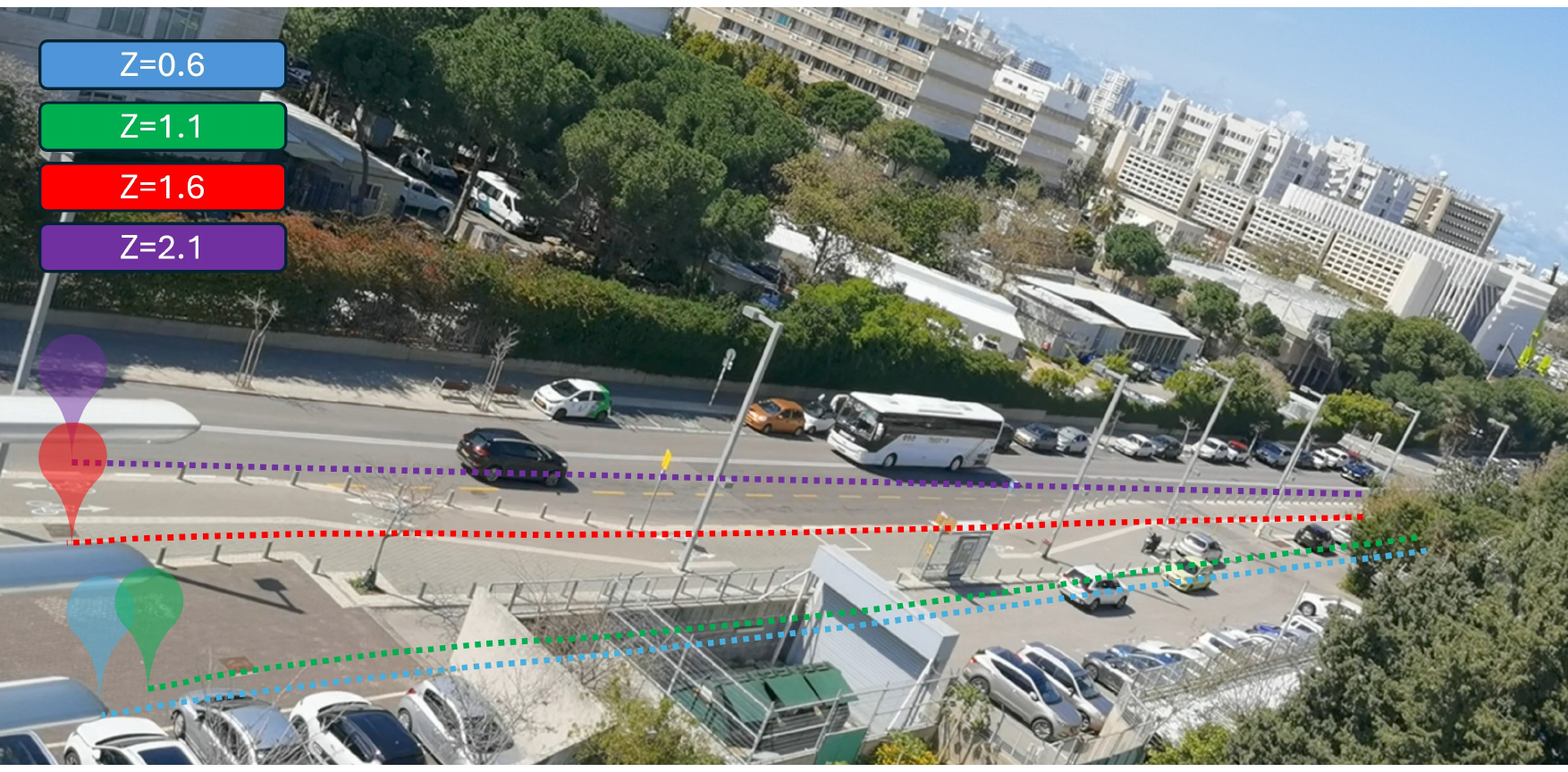}
\caption{Matched-filter–derived initializations at four fixed depths 
($Z=0.6,1.1,1.6,2.1$~m). 
For each depth, the observed $\sigma_\ell$ values are converted to $\rho_\ell=f^{-1}(\sigma_\ell;Z)$ under the monotonic near-field assumption.}
\label{fig:init_depth_sweep}
\end{figure}

To visualize the influence of depth on the initialization, Fig.~\ref{fig:init_depth_sweep} shows four candidate configurations generated for the same matched-filter responses but evaluated at fixed depths $Z \in \{0.6,1.1,1.6,2.1\}\,\mathrm{m}$, and with regularizing coefficients $\lambda_{\kappa}=\lambda_{\theta}=\lambda_{\ell}=1.0$.
Each line displays the corresponding preliminary loci 
$\{(x_\ell+\rho_\ell\cos\phi_\ell,\;y_\ell+\rho_\ell\sin\phi_\ell)\}$ prior to optimization. 
Depths exceeding $1.6$~m yield distorted near-field mappings. However, the monotonic restriction in~\eqref{eq:monotonic_rho} reduces discontinuities and promotes stable initialization for the optimization stage.

\begin{figure}[t]
\centering
\includegraphics[width=\linewidth]{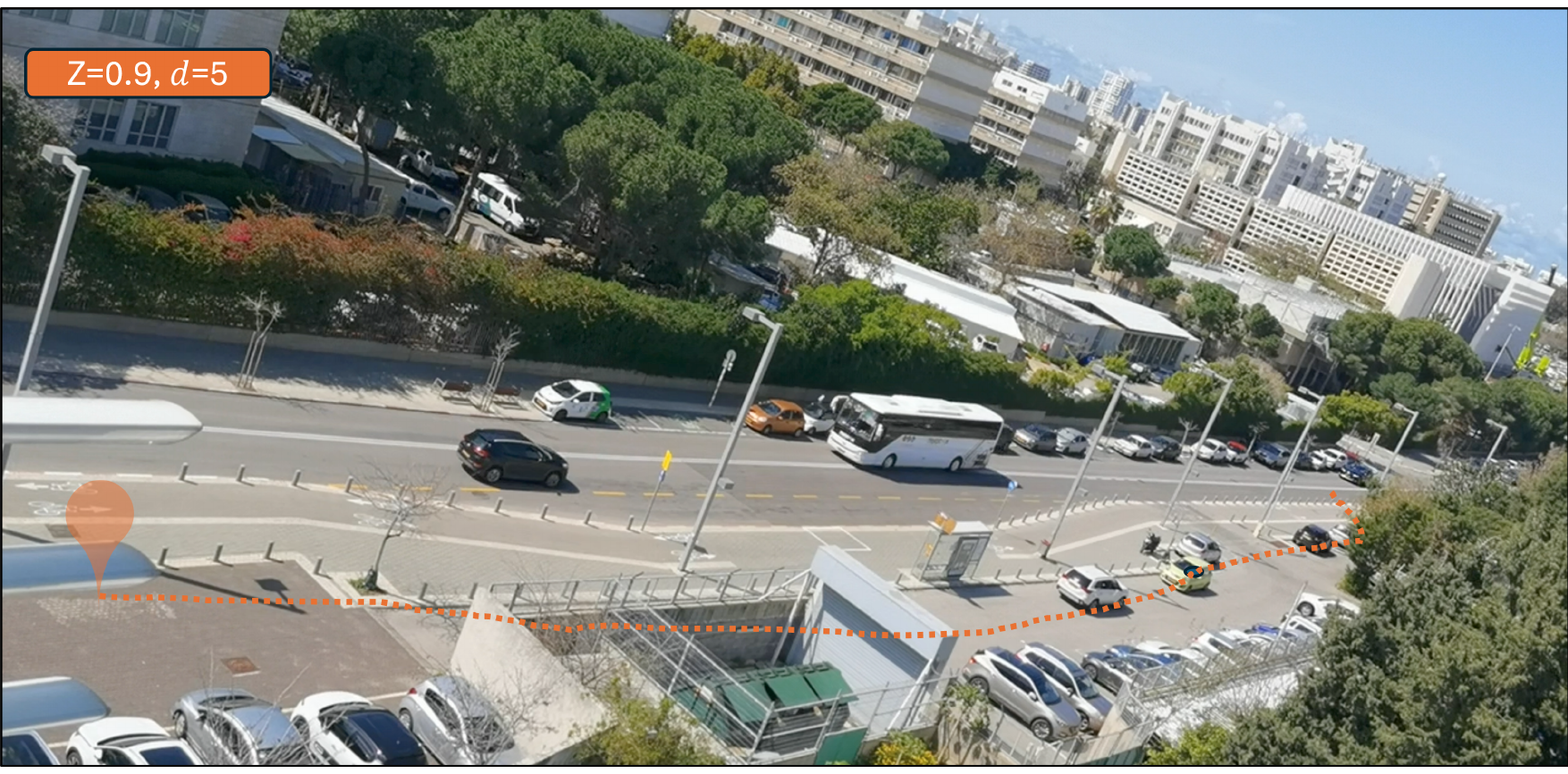}
\caption{High-order polynomial fit ($5^\text{th}$-degree) applied to 
the matched-filter initialization points $(x^*_\ell, y^*_\ell)$. 
While not metrically accurate, the fitted curve qualitatively captures large-scale geometry such as the fiber’s lateral crossing beneath the street.}
\label{fig:init_polyfit}
\end{figure}

Although the coupling-based initialization is not intended as a final estimate, 
fitting a higher-order polynomial to the matched-filter–derived $\rho_\ell$ values already reveals consistent geometric trends—most notably the lateral crossing of the fiber beneath the roadway (see Fig.~\ref{fig:init_polyfit}). 
In the example shown, a $5^\text{th}$-degree polynomial is used. Comparable crossing behavior persists for higher polynomial orders, confirming the trend observed in the GPS-based optimization in Fig.~\ref{fig:ExperimentStreet}.

\section{Sensitivity Analysis}
\label{app:sensitivityAnalysis}

\begin{figure*}
    \centering
    \includegraphics[width=0.7\linewidth]{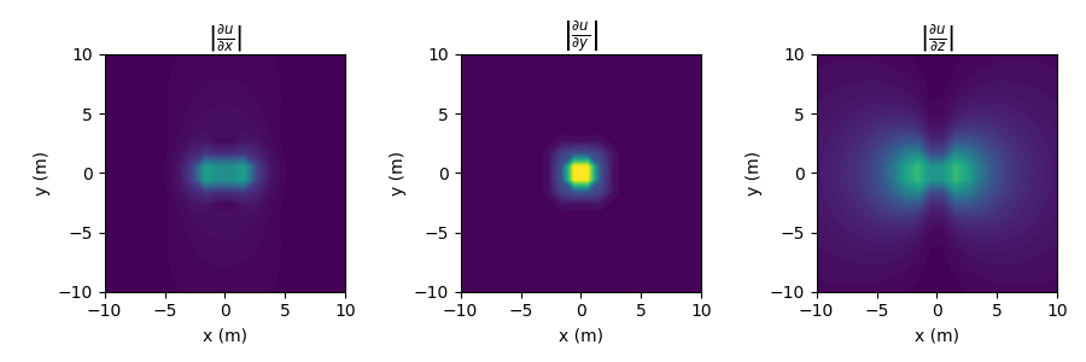}
    \caption{Sensitivity analysis at a depth of 1 m.}
    \label{fig:sensitivity}
\end{figure*}

Figure~\ref{fig:sensitivity} shows the sensitivity of strain derivatives with respect to spatial axes.

{
\subsection{MLP Ablation Study}
We justify the neural network (MLP) as a smooth reparameterization of the fiber trajectory by comparing it against direct optimization of the fiber-point coordinates. To compare fairly, both are initialized identically at the straight-line interpolation between the two fiber-edge anchors, and optimized with the same DAS data, regularizer, and learning rate. {Figure~\ref{fig:MLPAblation} shows that the MLP reaches the accuracy floor ($\approx$14~cm) within about 40 steps, whereas direct coordinate optimization requires roughly 200 steps to reach the same accuracy in this simplified scenario, corresponding to a difference of about a factor of 5 in convergence speed.}

\begin{figure}
    \centering
    \includegraphics[width=1\linewidth]{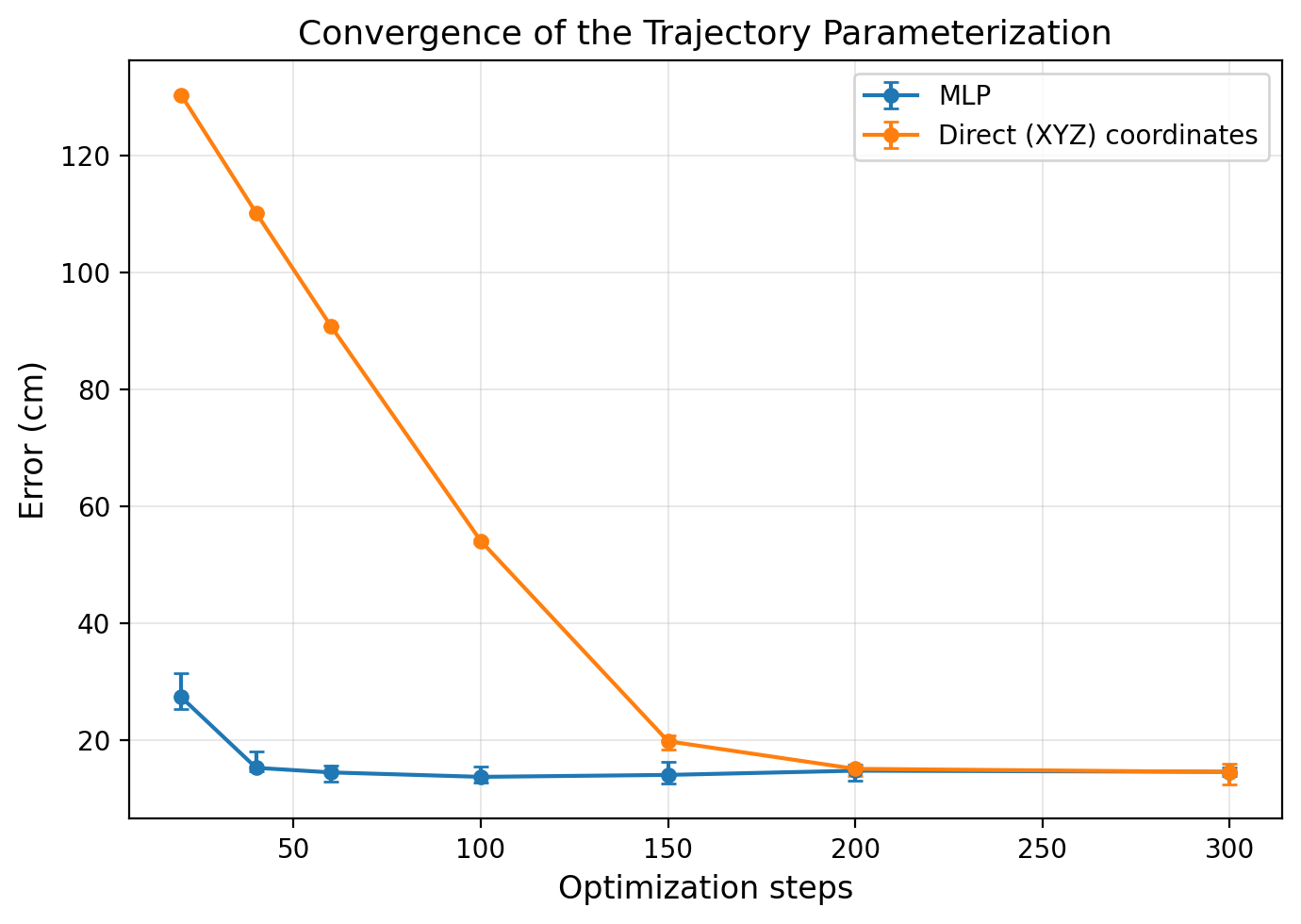}
    \caption{Convergence of the MLP vs.\ direct coordinate optimization (20 trials), both initialized at the same straight line between the fiber-edge anchors on a moderately-wiggly fiber. The MLP reaches the accuracy floor in $\sim$40 steps. Direct coordinate optimization needs $\sim$200 steps to reach the same accuracy.}
    \label{fig:MLPAblation}
\end{figure}



}

\section{Frenet-Serret equations}
\label{app:FernetSerret}

In modeling physically realistic optical fibers, it is useful to impose smoothness constraints on fiber geometry. Two geometric quantities that naturally characterize the local shape of a 3D curve are \textit{curvature} and \textit{torsion}. These are fundamental to the Frenet-Serret formalism and can be used to construct regularizers that penalize sharp bends and twists in the fiber path.

Let $\vec{S}(l)$ denote the spatial location of the fiber at arc-length parameter $l$. The unit tangent vector is defined as:
\begin{equation}
    \hat{T}(l) = \frac{d\vec{S}}{dl}  \ .
\end{equation}

The curvature $\kappa(l)$ quantifies the rate of change of the tangent vector and is given by:
\begin{equation}
    \kappa(l) = \left\| \frac{d\hat{T}}{dl} \right\| = \left\| \frac{d^2\vec{S}}{dl^2} \right\|  \ .
    \label{eq:curvature}
\end{equation}
Curvature is zero for a straight fiber and increases with tighter bends.

To define the torsion, we introduce the normal vector:
\begin{equation}
    \hat{N}(l) = \frac{1}{\kappa(l)} \frac{d\hat{T}}{dl}  \ ,
\end{equation}
and the binormal vector:
\begin{equation}
    \hat{B}(l) = \hat{T}(l) \times \hat{N}(l) \ .
\end{equation}

Torsion $\tau(l)$ measures how much the curve deviates from being planar, i.e., how much it twists in 3D. It is defined as:
\begin{equation}
    \tau(l) = -\frac{d\hat{B}}{dl} \cdot \hat{N}(l)  \ .
    \label{eq:torsion}
\end{equation}

These quantities can be computed numerically using finite differences over the discretized fiber points. In the optimization of fiber geometry, curvature and torsion can be regularized via a penalty term in the loss function:
\begin{equation}
    \mathcal{L}_{\text{curv+torsion}} = \lambda_\kappa \cdot \mathbf{E}[\kappa(l)^2] + \lambda_\tau \cdot \mathbf{E}[\tau(l)^2]  \ ,
\end{equation}
where $\lambda_\kappa$ and $\lambda_\tau$ are weighting factors controlling the strength of curvature and torsion regularization, respectively.

By minimizing this loss, we encourage the fiber to be globally smooth, avoiding sharp kinks or excessive twisting that are physically implausible in buried fiber installations. This helps stabilize optimization and improves interpretability and realism of simulated or learned fiber configurations.

\section{GPS data trajectory}
\label{app:gpsTraj}
\begin{figure}
    \centering
    \includegraphics[width=1.0\linewidth]{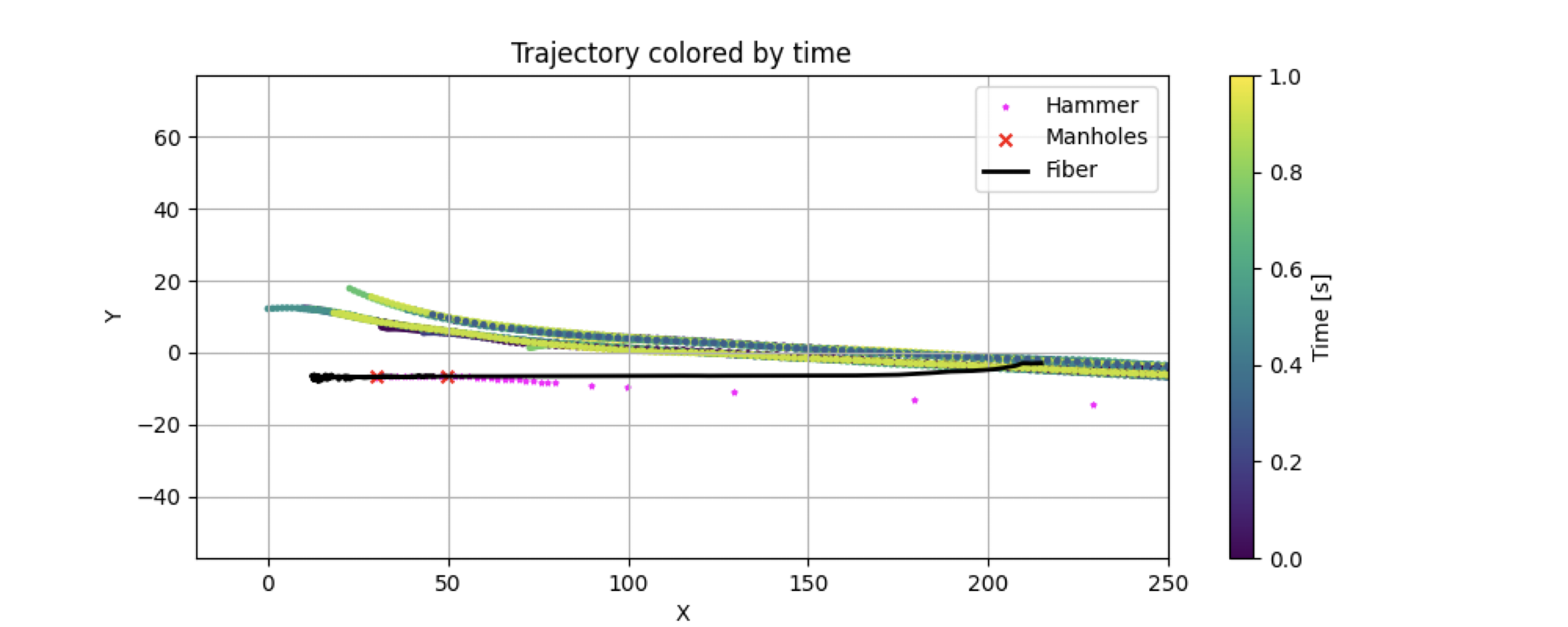}
    \caption{GPS trajectory of the moving vehicle, together with hammer-hit locations, manhole locations, and the optimized fiber.}
    \label{fig:gpsTraj}
\end{figure}

Figure~\ref{fig:gpsTraj} shows the optimized fiber location together with the GPS trajectory and hammer-hit locations from the experiment.

{
\section{Experimental error evaluation}
\label{app:expErrorEval}
In Figure \ref{fig:expErrorEval} we evaluate the fiber-localization error with respect to the ground truth as a function of the $x$ coordinate (the along-street direction). The bottom panel shows the lateral position of the ground-truth fiber (blue) together with the camera-based (green) and GPS-based (red) estimates. The top panel shows the corresponding local error, $|\mathrm{Camera}-\mathrm{GT}|$ and $|\mathrm{GPS}-\mathrm{GT}|$.

\begin{figure}
\centering
\includegraphics[width=1\linewidth]{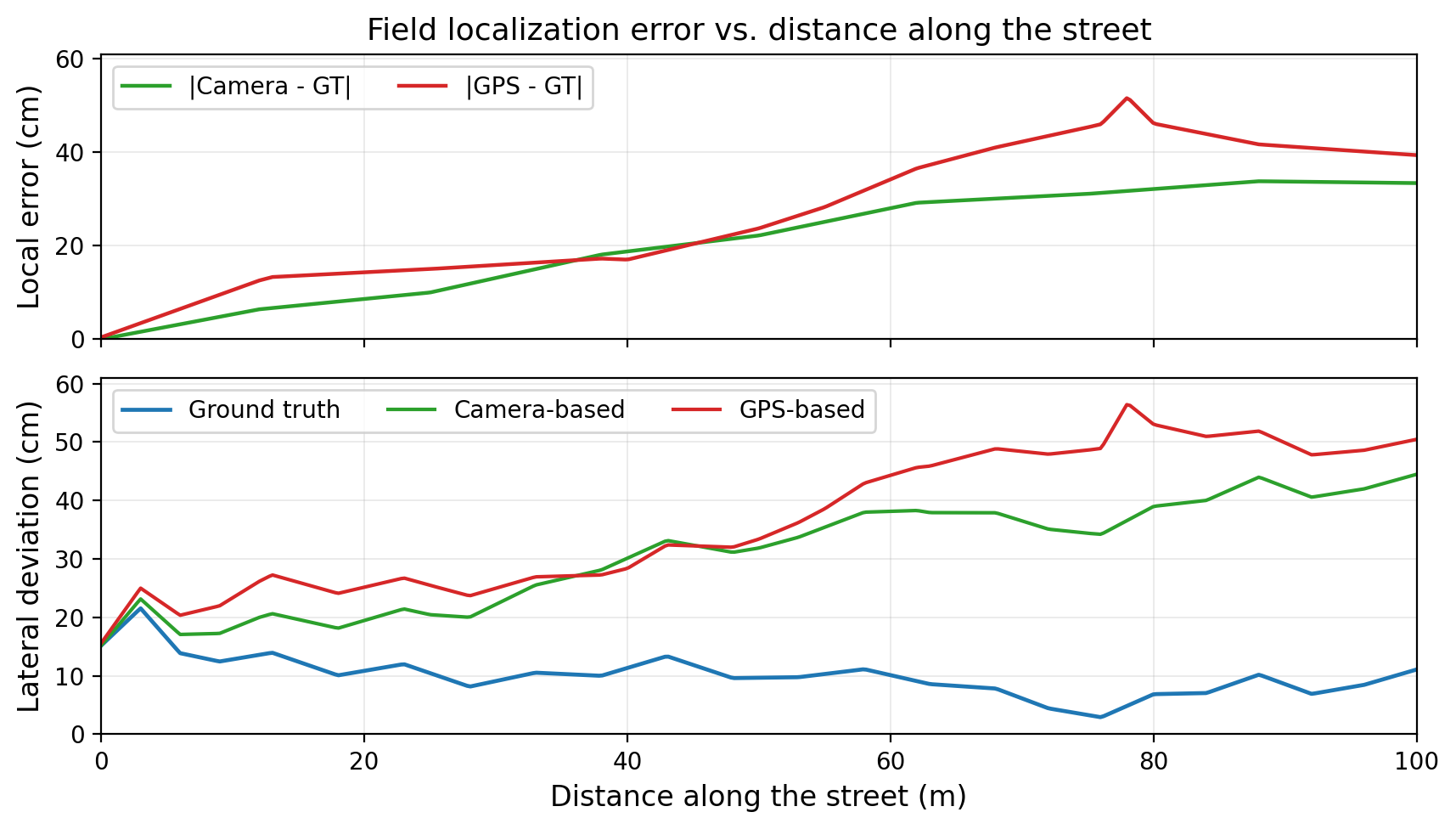}
\caption{Experimental fiber-localization error along the street ($x$ axis). Bottom: lateral ($y$) position of the ground-truth fiber (blue) and of the camera-based (green) and GPS-based (red) estimates. Top: local error of each estimate with respect to the ground truth, $|\mathrm{Camera}-\mathrm{GT}|$ and $|\mathrm{GPS}-\mathrm{GT}|$.}
\label{fig:expErrorEval}
\end{figure}
}

{
\begin{table}
\centering
\caption{Average per-pixel loss (Eq.~\eqref{eq:total_loss}) before and after optimization. Each row is averaged over the indicated number of sessions.}
\label{tab:mse_results}
\renewcommand{\arraystretch}{1.2}
\begin{tabular}{lcc}
\hline
\textbf{Session type (\# sessions)} & \textbf{Pre-opt. MSE} & \textbf{Post-opt. MSE} \\
\hline
Camera Experiment (11) & $7.90 \times 10^{-3}$ & $4.60 \times 10^{-3}$ \\
GPS Experiment (18) & $1.24 \times 10^{-2}$ & $1.16 \times 10^{-2}$ \\
Simulation (10) & $1.53 \times 10^{-3}$ & $1.49 \times 10^{-4}$ \\
\hline
\end{tabular}
\end{table}

Table~\ref{tab:mse_results} presents the average loss over sessions, where the
simulation conditions are similar to those in Section~IV. The reported values
are the average loss (over pixels) based on Eq.~\eqref{eq:total_loss}. The
variation between the results stems from the differences in the map sizes
($1800 \times 90$, $148 \times 170$ and $48 \times 100$, respectively) and in
the SNR. Despite the clear improvement from the pre- to the post-optimization
results, the loss differences are small since the car signal covers only a
small fraction of the pixels (about a single line within the whole map).

}

\section{YOLO model comparison and noise analysis}
\label{app:YOLOanalysis}
\subsection{Model Comparison}
We evaluated three YOLO models on the same experimental scene. Each model was run in tracking mode over a one-minute video segment containing two buses and several cars. For each model, we focused on the \textit{bus} and \textit{car} classes and compared tracking performance.

For each class, we extracted the following statistics:
\begin{itemize}
    \item The number of unique track IDs per class.
    \item The average number of frames during which a vehicle is continuously tracked.
    \item The average length of the longest continuous tracking fragment per vehicle.
\end{itemize}

\begin{figure}[t]
\centering
\includegraphics[width=1.0\linewidth]{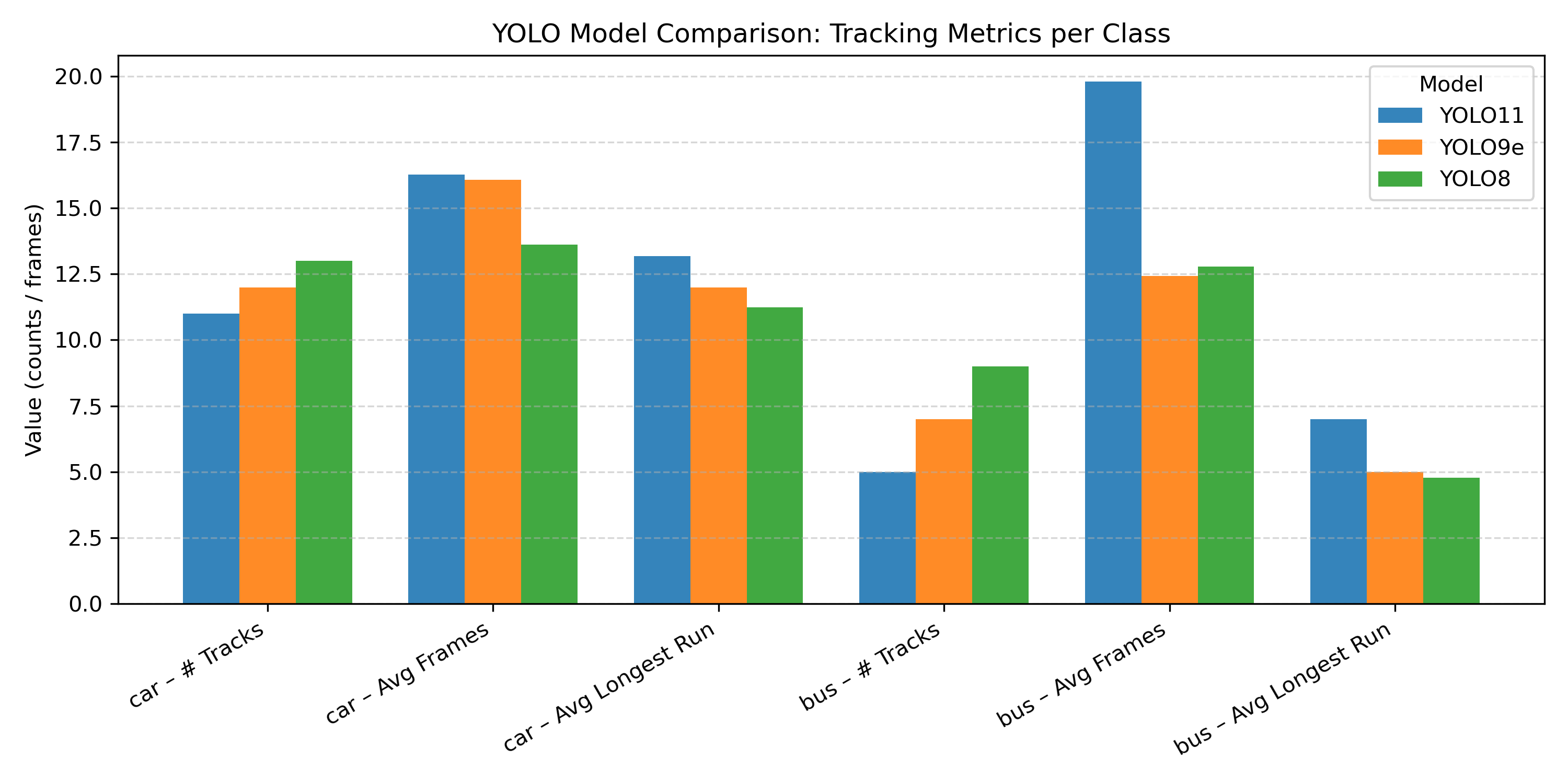}
\caption{Comparison of tracking performance across three YOLO models (YOLOv8, YOLOv9e, and YOLOv11) for the \textit{car} and \textit{bus} classes. 
Each group of bars represents a different tracking metric: the number of unique tracks, the average number of frames per track, and the average length of the longest continuous detection segment. 
YOLOv11 demonstrates the most stable tracking performance, particularly for the \textit{bus} class, with longer and more consistent track fragments.}
\label{fig:yolo-comparison}
\end{figure}

As shown in Fig.~\ref{fig:yolo-comparison}, \textbf{YOLOv11} detected five unique buses, while the other models produced more than five bus tracks in the same scene. YOLOv11 also maintained the longest continuous tracking fragments and the highest overall tracking duration (in frames).
Therefore, we conclude that YOLOv11 provides the most stable and consistent tracking performance among the tested models.

\subsection{YOLOv11 Spatial Localization Noise Analysis}

We evaluated the spatial noise distribution of YOLOv11 detections across the video frame. Using a one-minute sample containing two buses (one standard and one articulated), we extracted all bus-track data from this segment.

\begin{figure}[t]
    \centering
\includegraphics[width=1.0\linewidth]{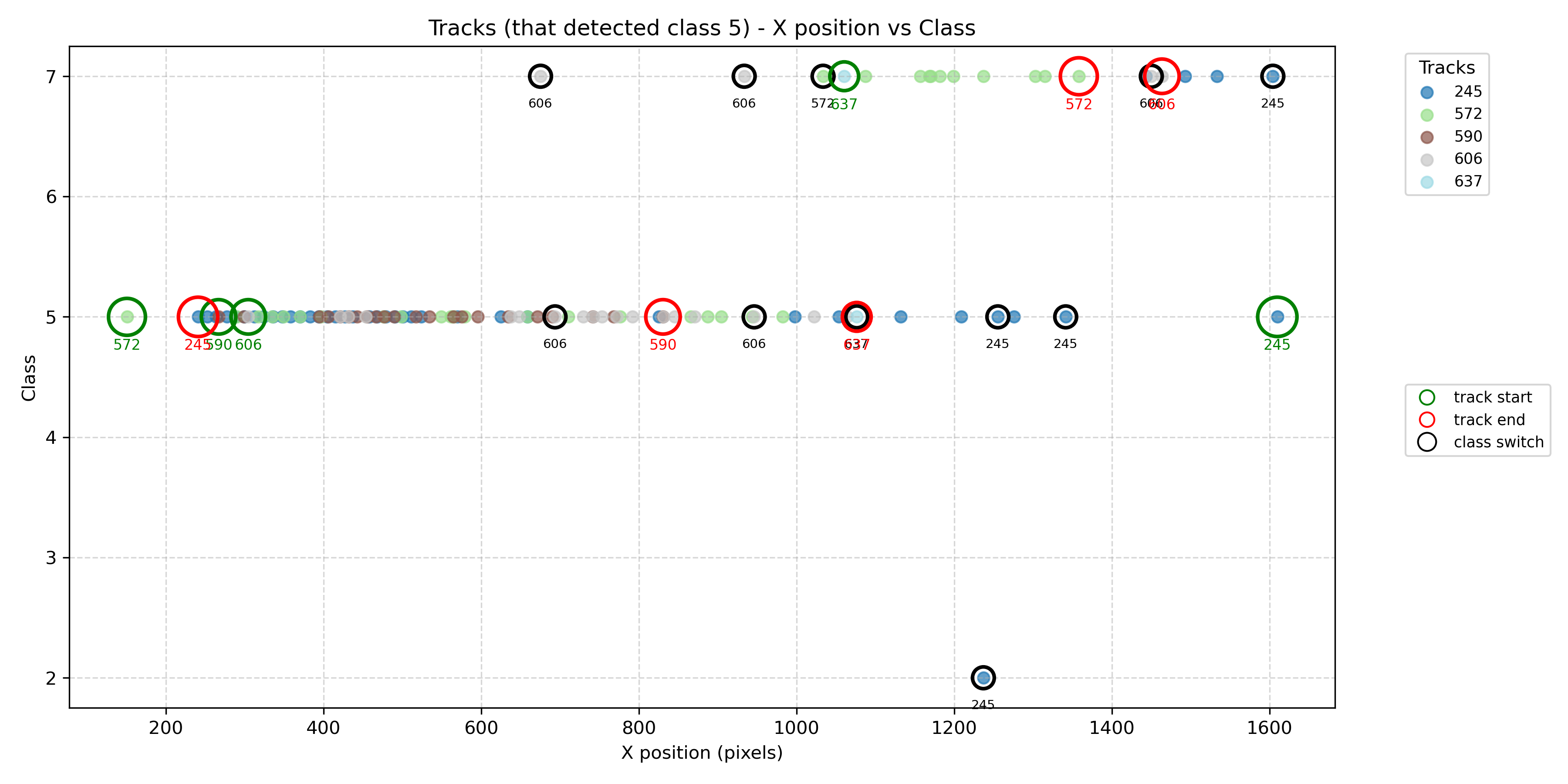}
\caption{YOLOv11 bus tracks as a function of horizontal frame position and detected class. 
Start and end points of each track are marked in green and red, respectively. Black markers denote class switch events where the detected class label changes between consecutive frames.}    \label{fig:tracks_x_position_vs_class}
\end{figure}

Figure~\ref{fig:tracks_x_position_vs_class} visualizes detections for each track as a function of horizontal frame position (\(X\)) and detected class. For each track, start and end points are marked in green and red, respectively, while black markers indicate class-switch events (label changes between consecutive frames). This plot highlights several patterns: the articulated bus is sometimes detected inconsistently (as two buses, as a bus and a truck, or as a single bus), and the right side of the frame exhibits more fragmented tracks and class switches than the left.

To quantify localization noise, each detection was projected from pixel coordinates into real-world meters using a homography transformation. For every track, we computed a 15-frame rolling mean and measured the per-frame deviation from this smoothed trajectory. The resulting noise magnitude captures the overall jitter of the detector in real-world space.

The analysis was performed twice: once before interpolation of the YOLO tracking data and once after interpolation. Interpolation is applied in two steps: (1) class interpolation, where each track is assigned the majority class label observed for that ID, and (2) frame interpolation, where missing detections within a track are filled using linear interpolation between neighboring frames to produce continuous trajectories.

For each version (prior and post interpolation), detections are grouped into 10 spatial bins along the horizontal axis, with $x = 0$ defined at the bus station in the middle of the scene. Within each bin, we compute the mean noise magnitude and its standard deviation. A dashed line at $35$ cm is included as a reference threshold. This value is based on related simulations (see Fig.~\ref{fig:heatmap_loc_noise}), where localization uncertainties exceeding $\approx 35$ cm caused trajectory optimization to degrade or fail. Thus, remaining below this threshold indicates that detections are sufficiently stable for reliable downstream processing.

\begin{figure}[t]
    \centering
    \includegraphics[width=\linewidth]{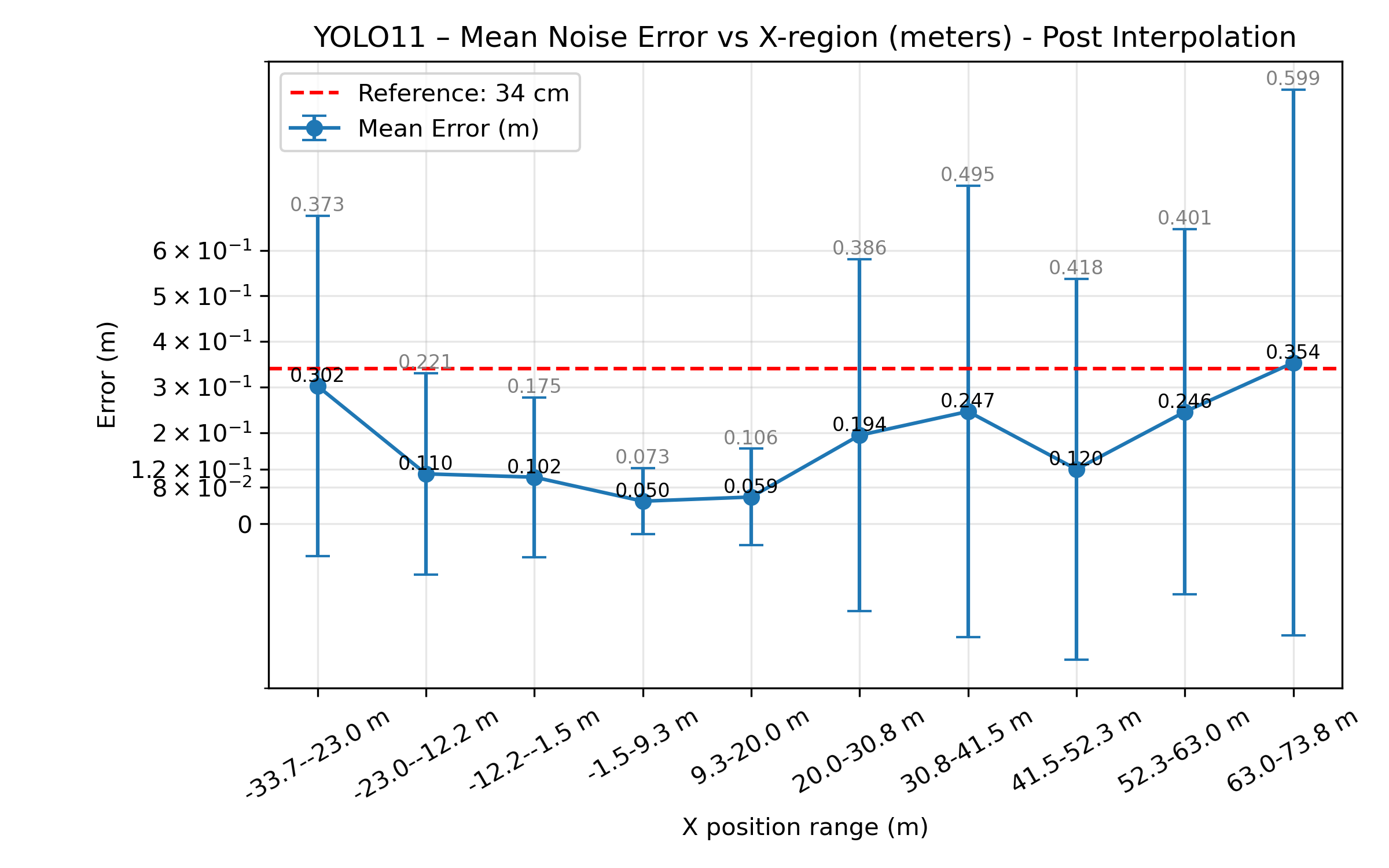}
    \caption{Mean localization noise as a function of horizontal position for YOLOv11 detections after interpolation. Error bars represent standard deviation of noise within each spatial bin. The red dashed line indicates a 35 cm reference threshold derived from simulation-based optimization sensitivity.}
    \label{fig:yolo-noise-post}
\end{figure}

\begin{figure}[t]
    \centering
    \includegraphics[width=\linewidth]{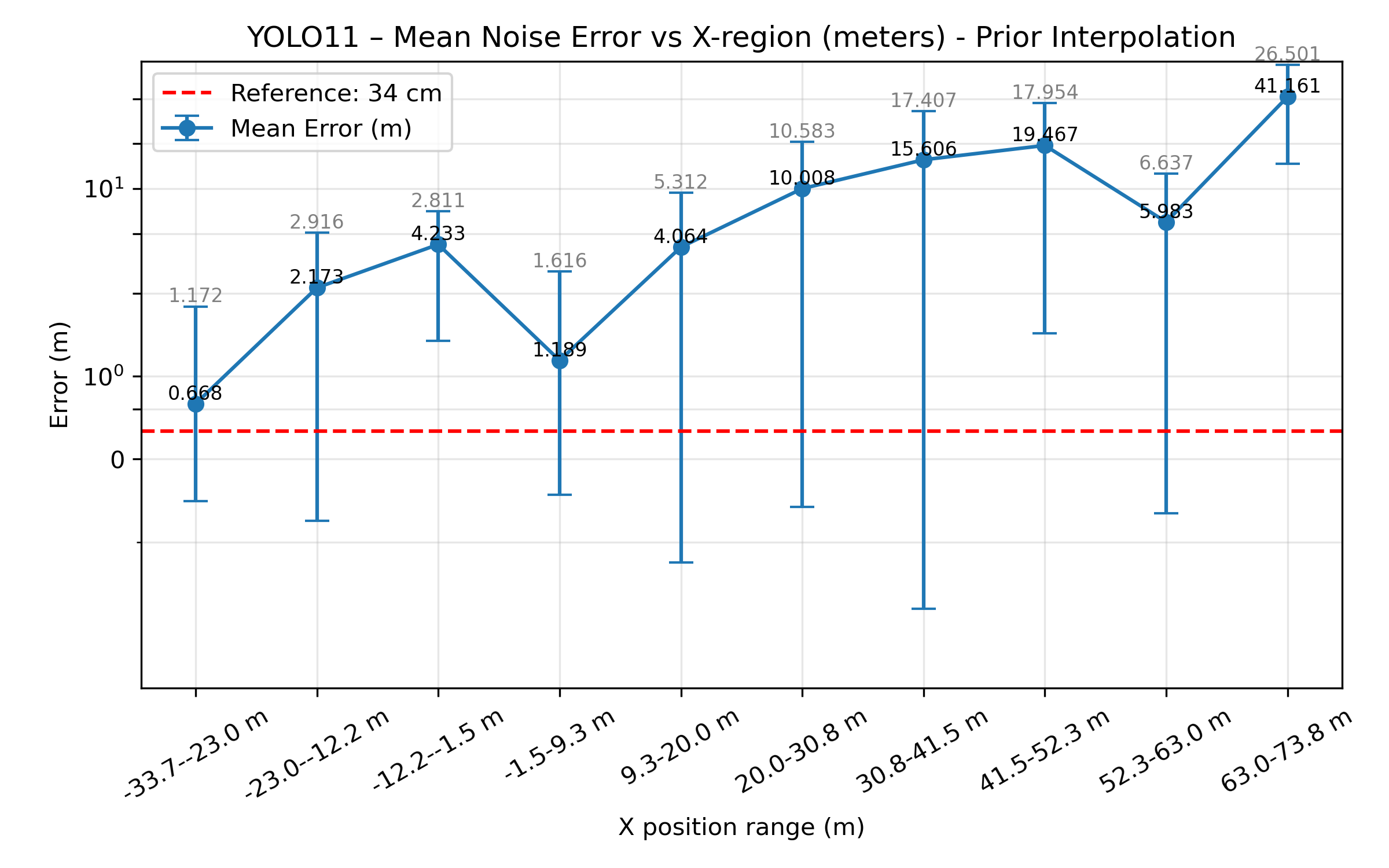}
    \caption{Mean localization noise as a function of horizontal position for YOLOv11 detections before interpolation. Noise increases significantly toward the scene periphery due to missing detections and unstable trajectories.}
    \label{fig:yolo-noise-prior}
\end{figure}

Figure~\ref{fig:yolo-noise-post} shows that interpolation substantially reduces YOLOv11 localization noise. Noise remains below the $35$ cm reference threshold near the station center, reaching a minimum of $5-7$ cm in the range $-1.5\,\text{m} \le x \le 9.3\,\text{m}$. Toward the field-of-view edges, error increases progressively, with mean values of approximately $0.35-0.60$ m, especially in the far-right bins. Before interpolation, localization noise increases sharply with distance from the camera center and exceeds 15 m in some outer regions (Fig.~\ref{fig:yolo-noise-prior}). This confirms that interpolation is critical for stable track continuity and spatial accuracy. Across all spatial bins, post-interpolation errors are one to two orders of magnitude lower than raw YOLO output.

\section{Regularizers}
\label{app:regularizers}

\paragraph{Curvature and torsion regularization}  
The local curvature \(\kappa_n\) and torsion \(\tau_n\) of the fiber are computed using the Frenet-Serret formulas (see Appendix~\ref{app:FernetSerret} for full definitions). Large values of \(\kappa_n\) or \(\tau_n\) correspond to non-physical sharp bends or twists. We penalize total curvature and torsion along the fiber via:
\begin{gather}
\label{eq:kappa}
    R_{\kappa} = \frac{1}{N-2} \sum_{n=2}^{N-1} \left( \kappa_n^2 + \tau_n^2 \right) \ ,
\end{gather}
which encourages gently varying, physically realistic trajectories.

\paragraph{Anchor proximity prior}  
We denote by \textit{anchors} any prior knowledge of fiber location (e.g., manholes). Anchors are used both in initialization and in optimization regularization. Given known fixed anchor points \(\{\mathbf{A}_m\}_{m=1}^M\) (e.g., from field measurements or design constraints), we encourage the reconstructed fiber to cross them. The penalty term for not doing so is:
\begin{gather}
    R_a = \frac{1}{M} \sum_{m=1}^M \min_{1 \le n \le N} \left\| \mathbf{S}_n - \mathbf{A}_m \right\|^2 \ .
\end{gather}

We found that \(R_{\kappa}\) was less suitable for the neural-network-based optimization, where it introduced numerical instability and was difficult to tune, but it did provide useful guidance during the matched-filter-based determination of the \(y\)-offset.

{
\section{Anchors Uncertainty Analysis}
\label{app:anchorsUncertainty}
Figure \ref{fig:anchorsUncertainty} shows a comparison between the Hard and Soft anchor cases. The conditions are similar to those in Section \ref{sec:simulations}. In addition, a positional error of increasing magnitude (a fixed displacement in a random horizontal direction) was added to each of the two anchors.

\begin{figure}
    \centering
    \includegraphics[width=1.0\linewidth]{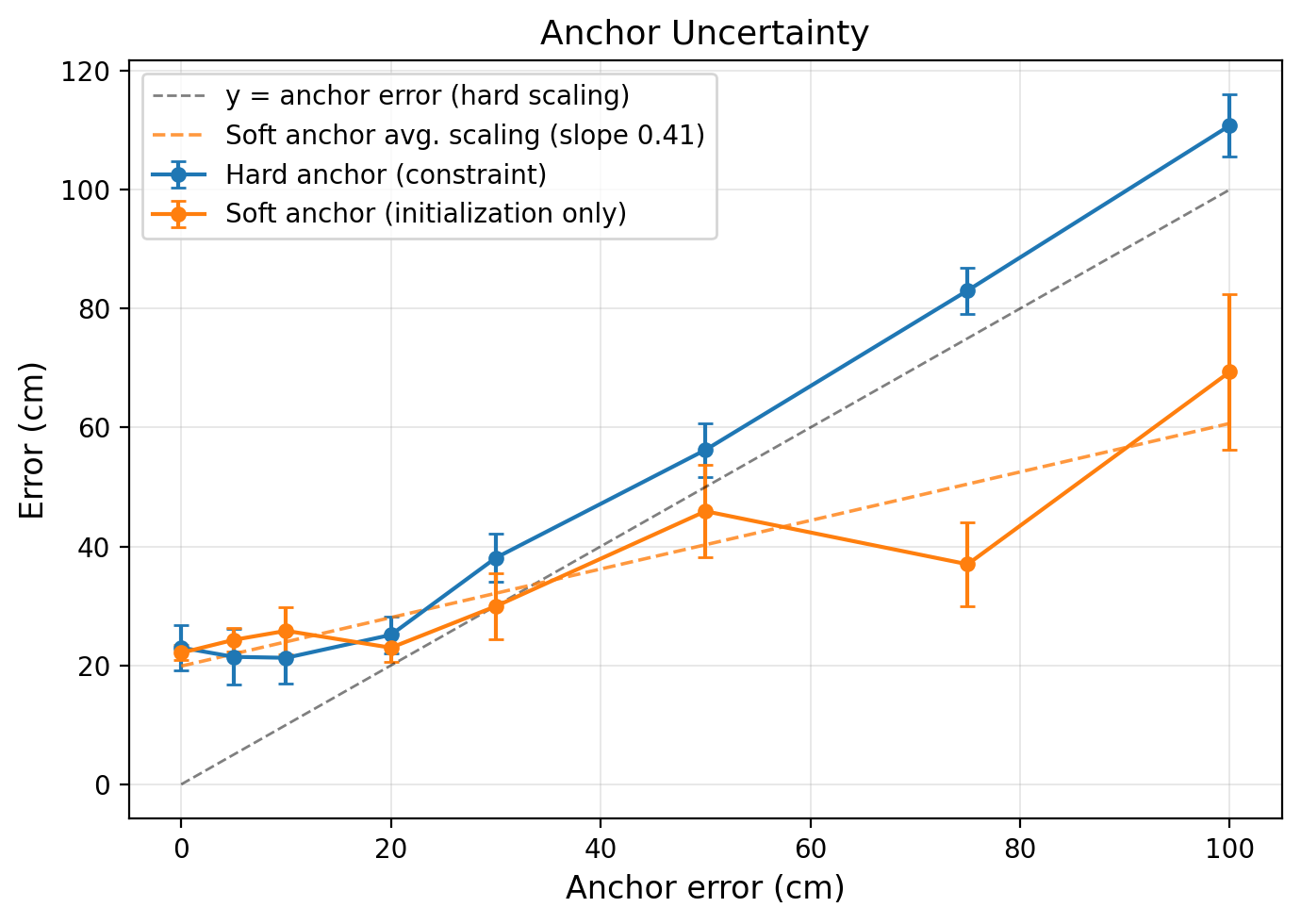}
    \caption{Effect of anchor uncertainty on localization error, and comparison between Hard and Soft anchors. Results are reported as mean $\pm$ standard deviation over 40 trials. Accuracy decreases as anchor errors increase, while the scaling differs between the Hard and Soft anchor cases. The fiber was buried at a depth of $90$ cm, with an average horizontal offset of $3.75$ m.}
    \label{fig:anchorsUncertainty}
\end{figure}

}

{
\section{Point-load source Approximation}
\label{app:PLSapprox}

As mentioned in the text, we approximate the DAS response of the moving vehicle by a single point source. This approximation is accurate for typical light vehicles but may break down for heavy, multi-axle ones. In Figure \ref{fig:PLSapprox} we examine the optimization performance for three vehicles of increasing mass and axle count - 1.6 t (4 wheels), 8 t (4 wheels), and 24 t (8 wheels), each at two burial depths (0.9 m and 0.3 m). Other conditions follow Section \ref{sec:simulations}. The matched multi-wheel inversion (blue) remains sub-meter in every case (median 11-24 cm), whereas the point-load approximation (orange) degrades systematically with vehicle mass and axle count: median error grows from $\approx$45 cm (light car) to $\approx$70 cm (mid truck) to $\approx$215 cm (heavy truck). Burial depth has only a minor effect at the 10 m gauge length used here. The approximation is therefore adequate for typical passenger vehicles but should be replaced by the multi-wheel model for heavy or multi-axle vehicles.

\begin{figure}
    \centering
    \includegraphics[width=1\linewidth]{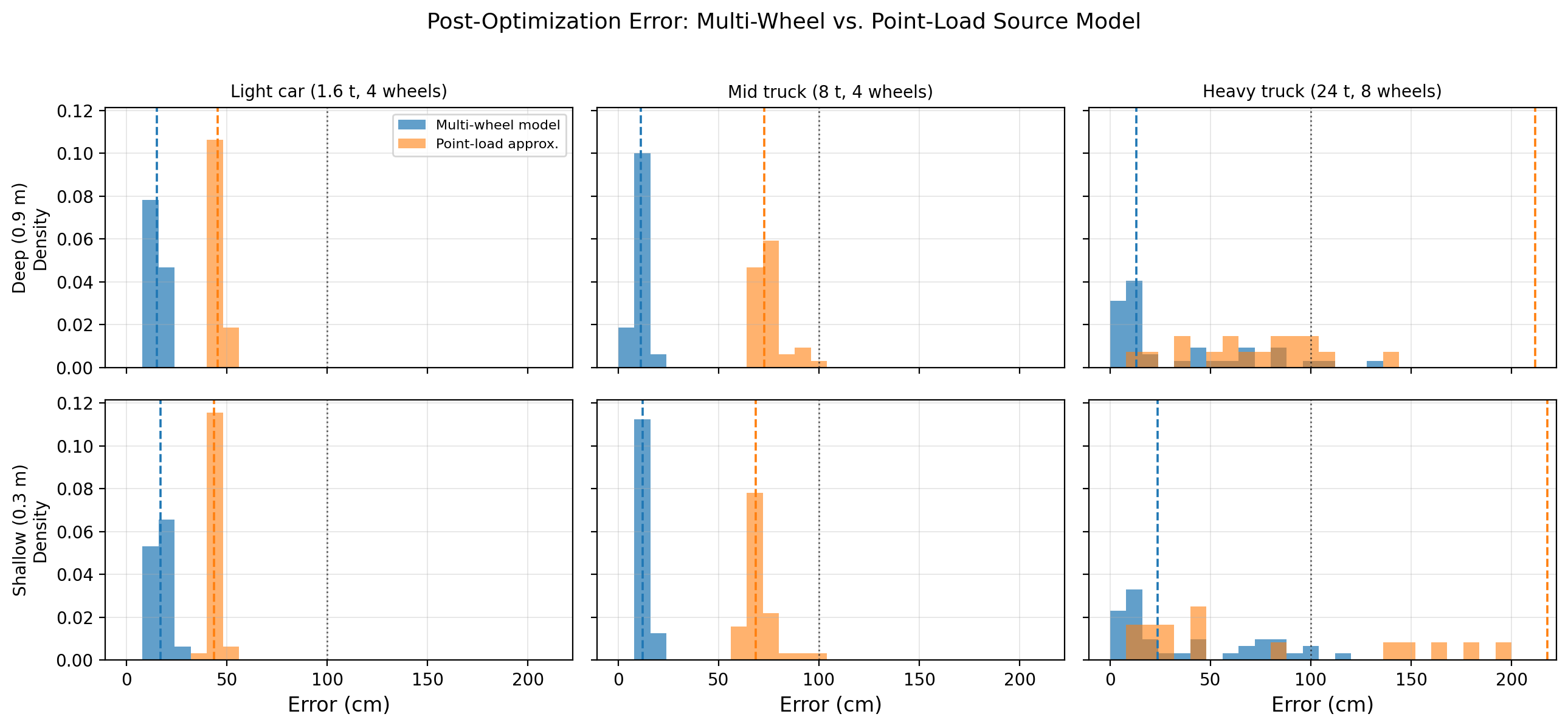}
    \caption{Post-optimization error distributions over 40 trials for the multi-wheel model (blue) and the point-load approximation (orange), over 6 cases: light, mid, heavy vehicle (columns), and deep, shallow (rows).}
    \label{fig:PLSapprox}
\end{figure}

}

{
\section{Fiber Scalability}
\label{app:FiberScalability}
Our suggested method was validated for a 100 m segment. However, we argue that it can be generalized by dividing any longer fiber into shorter segments and stitching them together. One way to connect the fiber estimations is by introducing a hard anchor at the edge of each 100 m fiber segment. In Figure \ref{fig:FiberScalability} we show a case where a 1 km fiber was divided into 10 segments of 100 m, each optimized independently. Here, we also added another constraint: we required not only joint anchors between segments (to ensure a continuous trajectory), but also equality of the first derivative along the fiber (to ensure a smooth fiber trajectory).

This technique works well even when inhomogeneities exist along the fiber, as seen in the bottom row, where the Poisson ratio and burial depth vary between segments and are unknown to the optimizer: the Poisson ratio varies randomly between 0.23 and 0.27, and the burial depth varies by $\pm5$ cm around the nominal 90 cm.

\begin{figure}
    \centering
    \includegraphics[width=1\linewidth]{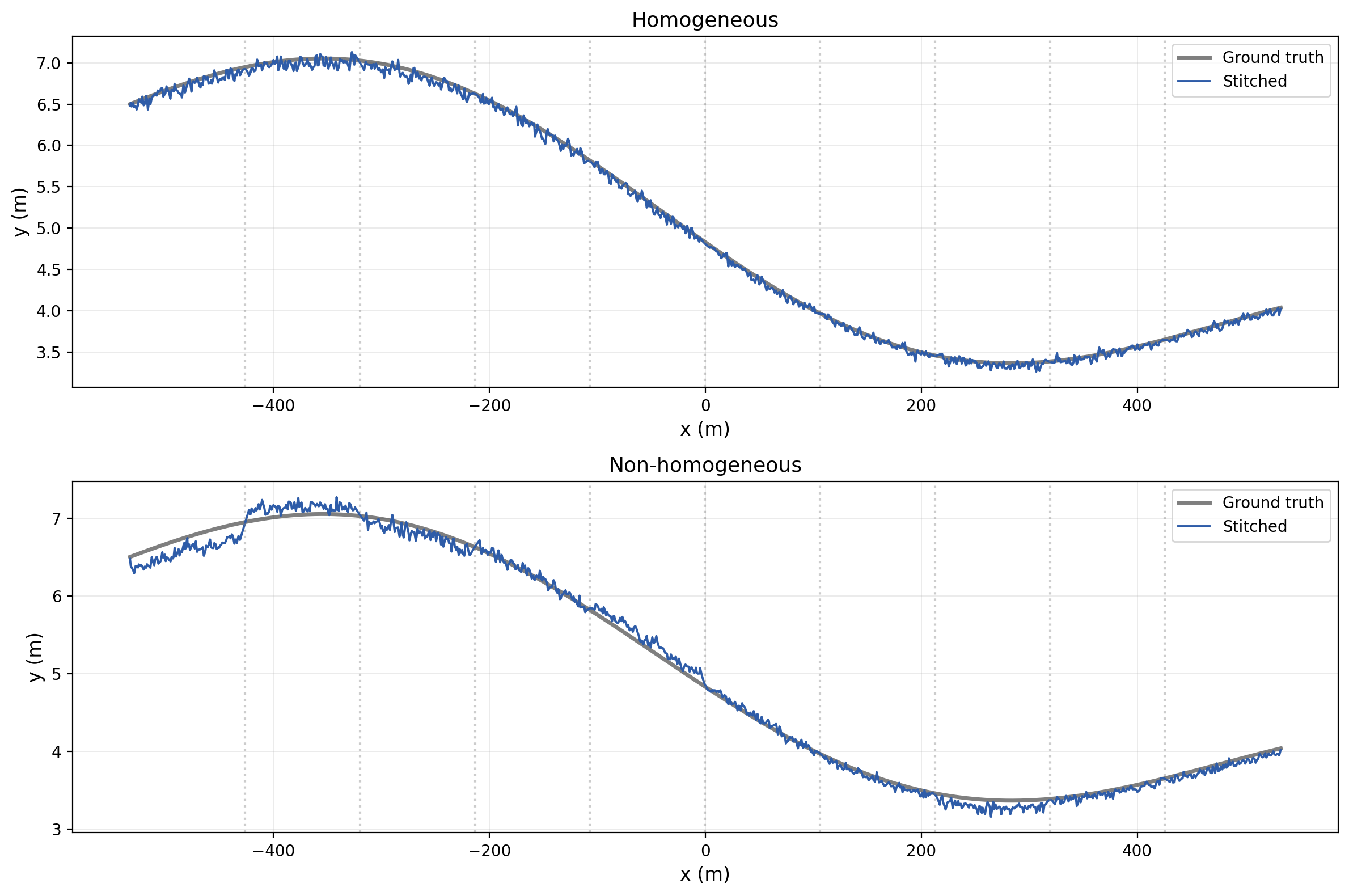}
    \caption{1 km stitched reconstruction (blue) vs. ground truth (gray), dotted lines mark the 100 m joints. Top: homogeneous conditions. Bottom: non-homogeneous medium (different Poisson ratio and fiber burial depth).}
    \label{fig:FiberScalability}
\end{figure}

}

{
\section{Methods Comparison}
\label{app:MethodsComparison}
{Here we compare our method (Ours) against three baseline approaches: spline interpolation (Spline), peak-channel tracking (Peak), and matched-filter initialization (Matched). The scenario is similar to Section~\ref{sec:simulations}, with two anchors at the fiber edges. Three use-cases were evaluated, ranging from low to high fiber curvature and from high to low SNR: easy - a straight fiber, SNR $=100$, medium - a sinusoidal fiber of amplitude $1.5$ m (one period), SNR $=40$, GPS noise $3$ cm, and hard - amplitude $2.0$ m ($1.5$ periods), SNR $=30$, GPS noise $5$ cm.} Figure \ref{fig:MethodsComparison} shows that for easy conditions the spline interpolation is even more precise than our method, whereas for any more complicated scenario our method has a significant advantage. The peak- and matched-filter baselines are limited in all cases because they cannot recover the fiber's lateral offset from the vehicle path.

\begin{figure}
    \centering
    \includegraphics[width=0.9\linewidth]{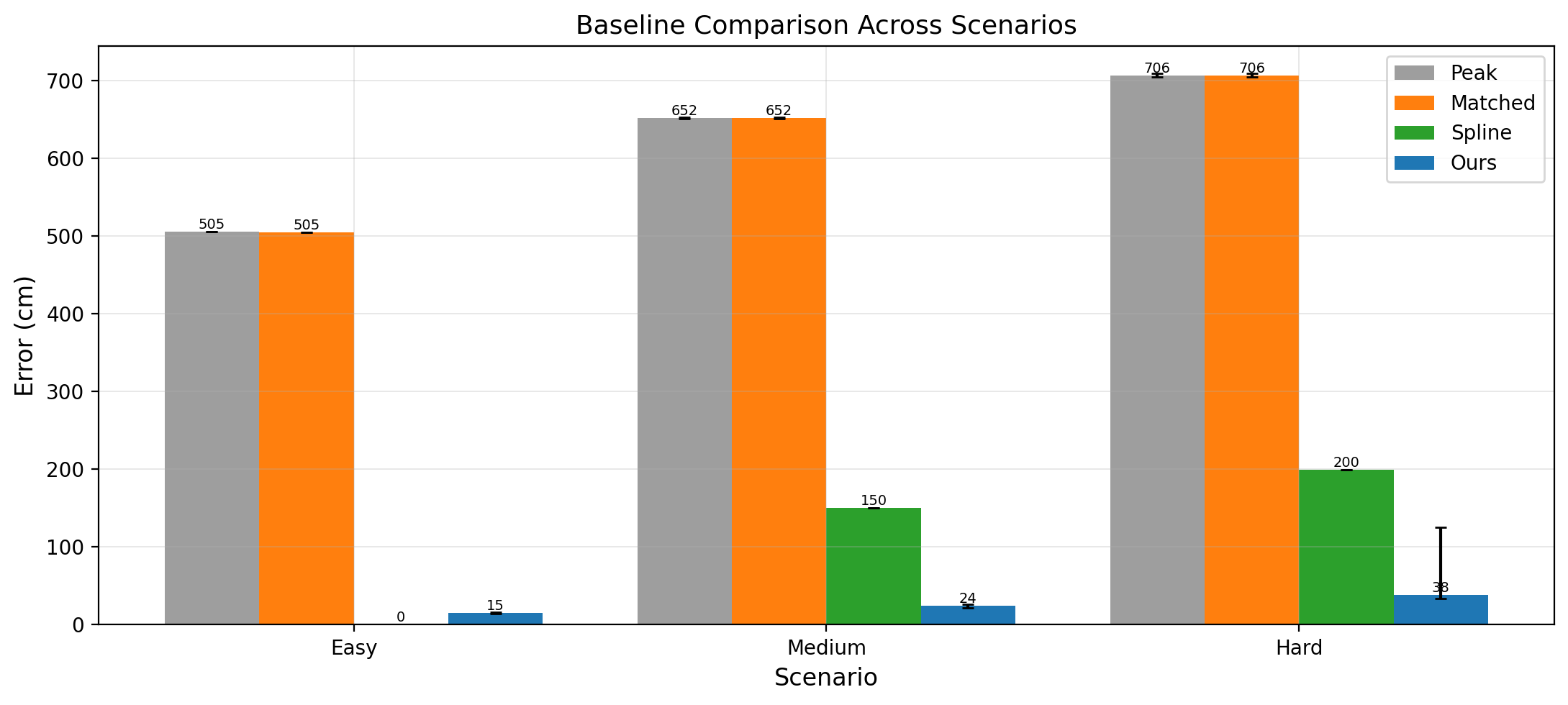}
    \caption{Baselines across scenarios, averaged over 20 {trials}.}
    \label{fig:MethodsComparison}
\end{figure}
}


\end{document}